 \pdfoutput=1
 
\documentclass[10pt,journal,  oneside, twocolumn]{IEEEtran}

\usepackage{multirow}
\usepackage{latexsym}
\usepackage{graphicx}
\usepackage{float}
\usepackage{amsmath}
\usepackage{amsthm}
\usepackage{lipsum}
\usepackage{subfig}
\usepackage{graphicx}
\usepackage{authblk}
\usepackage{bm}
\usepackage{amsthm}

\usepackage{colortbl}

\newtheorem*{remark}{Remark}

\makeatletter  
\newif\if@restonecol  
\makeatother

\usepackage[linesnumbered,ruled,vlined]{algorithm2e}
\usepackage{algpseudocode}  
\usepackage{amsmath}

\usepackage{amssymb}

\usepackage{mathrsfs}
\usepackage{subfig}
\usepackage{caption}
\begin{document}

\title{Intelligent Task Offloading for Heterogeneous V2X Communications}


\author{Kai Xiong, Supeng Leng,~\IEEEmembership{Member,~IEEE},  
Chongwen Huang, Chau Yuen,~\IEEEmembership{Senior Member,~IEEE} and Yong Liang Guan,~\IEEEmembership{Senior Member,~IEEE}

\thanks{

K. Xiong and S. Leng are with School of Information and Communication Engineering, University of Electronic Science and Technology of China, Chengdu, 611731, China.}

\thanks{C. Huang and C. Yuen is with Singapore University of Technology and Design (SUTD), Singapore.}

\thanks{Y. L. Guan is with the School of Electrical and Electronic Engineering, Nanyang Technological University, Singapore.}


\thanks{This work was partly supported by National Key R\&D Program of China (No.2018YFE0117500), the Science and Technology Program of Sichuan Province, China (No.2019YFH0007), and the EU H2020 Project COSAFE (MSCA-RISE-2018-824019).}

\thanks{The corresponding author is Supeng Leng, email: spleng@uestc.edu.cn}

}


\maketitle


\begin{abstract}

With the rapid development of autonomous driving technologies, it becomes difficult to reconcile the conflict between ever-increasing demands for high process rate in the intelligent automotive tasks and resource-constrained on-board processors. Fortunately, vehicular edge computing (VEC) has been proposed to meet the pressing resource demands. Due to the delay-sensitive traits of automotive tasks, only a heterogeneous vehicular network with multiple access technologies may be able to handle these demanding challenges. In this paper, we propose an intelligent task offloading framework in heterogeneous vehicular networks with three Vehicle-to-Everything (V2X) communication technologies, namely Dedicated Short Range Communication (DSRC), cellular-based V2X (C-V2X) communication, and millimeter wave (mmWave) communication. Based on stochastic network calculus, this paper firstly derives the delay upper bound of different offloading technologies with a certain failure probability. Moreover, we propose a federated Q-learning method that optimally utilizes the available resources to minimize the communication/computing budgets and the offloading failure probabilities. Simulation results indicate that our proposed algorithm can significantly outperform the existing algorithms in terms of offloading failure probability and resource cost.

 
\end{abstract}

\begin{IEEEkeywords}
Vehicular Edge Computing, DSRC, C-V2X, mmWave, Federated Q-Learning.

\end{IEEEkeywords}

\IEEEpeerreviewmaketitle






\section{Introduction}

\IEEEPARstart {M}{odern} transportation systems have evolved vehicles with the automotive artificial intelligence which can make decisions through on-board processors and shares road information by Vehicle-to-Everything (V2X) communications. The emergence of services like intelligent cruise scheduling, road traffic management, and cooperative driving with the shared on-board computing resources, has prompted the need for significant computing capacity and rigorous delay tolerance \cite{Zhang7907225}. Previous literature \cite{Zhou8535091} remarked that an autonomous vehicle can create $1$ GB of data per second from the on-board processor. However, the volume of the data will exponentially grow with the increasing number of autonomous vehicles. Without enough communication and computing resources, road risk estimation becomes leggy and inaccurate that may incur life-threatening problems.

Due to the constrained on-board processing power and communication bandwidth in vehicular networks, it is impractical to raise the computing and communication capacity of an individual vehicle for automotive applications. To cope with the scarce computing resources problem, Vehicular Edge Computing (VEC) technology has been proposed to support the computing-thirsty tasks in the intelligent transportation systems, in which the computing capabilities of vehicles can be improved through offloading the overburden tasks of a vehicle to the adjacent vehicles or edge servers \cite{Qiao8879573}. However, previous work \cite{{Dai8493149Dai},{Chai8918338}} mainly pays attention to the task offloading schemes with sufficient communication resources. However, the performance of VEC is significantly constrained by the limited communication bandwidth in a practical vehicular network.



On the other hand, the use of heterogeneous V2X communications has great potential to enhance the communication capability for the large scale application of autonomous vehicles \cite{Chen8784150}. Vehicle-to-Vehicle (V2V) communications are the major transmission form in the platoon-based edge computing paradigm, where a platoon of vehicles with sufficient on-board computing resources and communication bandwidth can offer additional mobile computing resources cooperatively \cite{{Qiao8436044}}. Specifically, there are three widely used types of V2V communication technologies, including the Dedicated Short Range Communications (DSRC) communication, cellular-based V2V (C-V2V) communication, and mmWave communication. As for the Vehicle-to-Infrastructure (V2I) communication, it can support the infrastructure-based edge computing \cite{{Zhang8403956}}, in which the overburden task is offloaded from vehicles to the nearby base station that provides the communication service. In this paper, we investigate the cellular-based V2I communication (C-V2I) that requires vehicles to stay within cellular coverage. In addition, DSRC and mmWave V2V operate on the license free bands \cite{{Mavromatis8904214},{Ghasempour8088544}}. However, the C-V2V and C-V2I work on the licensed band to provide paid communication service.

In the context of VEC with heterogeneous V2X connections and computing servers, the offloading reliability is tightly affected by the selection of access technologies and offloading targets. Malfunctions of any part (transmission may fail or task processing may be interrupted) deteriorate the VEC performance. However, only a few work has investigated the integrating failure probability of computing and communication processes in a heterogeneous VEC network, or incorporated this failure probability into the VEC optimization objective. Therefore, with the constrains of the offloading failure probability and resource cost, it is still an open challenge to attain the reliable and economical VEC services through the optimal selection of access technologies and offloading targets.

To fulfill these research gaps, our study exploits the heterogeneous VEC to optimize tasks offloading with multiple V2X technologies. First, we derive the upper bound of the offloading delay using different access technologies with variable failure probability. Based on the analysis of the upper bounds, a federated learning-based intelligent offloading scheme is proposed to minimize the failure probability and the resource cost in terms of communication cost and computing cost. Our federated Q-learning algorithm can parallelly exploit and share the local knowledge among vehicles. The main contributions are summarized as follows:





  

\begin{itemize}

\item We derive the upper bound of offloading delay for different V2X technologies by leveraging stochastic network calculus. It is the first closed-form of the upper bound for the offloading delay in the VEC. These upper bounds can guide the design of optimal offloading scheduling and the selection of communication forms for the task offloading.

\item We propose a new optimization model taking account of the communication and computing budgets as well as the failure probability. This is the first work to consider the offloading failure probability in the heterogeneous VEC. Simulations show that the cellular-based V2X (C-V2X) communication have the best performance in terms of the failure probability under different traffic loads. 

\item We design a federated learning-based parallel scheme with high scalability and fast convergence. The optimization model is decoupled and parallelly trained by a set of local Q-learning processes, which can accumulate global knowledge by exploiting different local action-state spaces, simultaneously. The proposed consensus Q-Table can amplify the knowledge sharing, and avoid the heavy communication overhead of the training phase. 




\end{itemize}

The remainder of this paper is organized as follows. Section II presents the related works. Section III introduces the system model. Section IV provides the platoon-based edge computing and infrastructure-based edge computing formations. Section V presents the optimization model and our two solutions on the V2X offloading selection. Section VI demonstrates the simulation results and the performance discussion. Finally, we draw the conclusion in Section VII.

\section{Related Work}

Existing work on VEC focuses on selecting the optimal offloading targets taking account of resource constraints. Dai \textit{et al.} \cite{Dai8493149Dai} investigated the multiple vehicles task offloading problem of VEC and proposed the two-step iterative algorithm to optimize offloading ratio and computation resource. Zhang \textit{et al.} \cite{Zhang7997360} proposed a Stackelberg game approach for the offloading candidates selection to optimize the utilities of both vehicles and VEC servers. Zhang \textit{et al.} \cite{Zhang8403956} regarded vehicles as caching servers and proposed a caching service migration scheme, where the communication, computing, and caching resources at the wireless network edge are jointly scheduled. However, these VEC solutions ignored the potential selection problem of optimal access technologies for task offloading.

Furthermore, each V2X technology has its own limitations. DSRC is a typical competition-based communication that is not suitable for multiple tasks communication due to scalability issues \cite{Sial8859331}. The C-V2X technology eliminates the above drawbacks, while it charges the fee to vehicles \cite{Zheng7293220}. In contrast, the mmWave communication works at the high-frequency unlicensed bands with the large available bandwidth \cite{Chen8784150}. However, it needs antenna beam alignment that incurs additional link budget \cite{{chogwentsp2019}}. Consequently, the academic society has been exploited a heterogeneous V2X architecture that complements shortcomings of individual V2X technology. Abboud \textit{et al.} \cite{Zhuang7513432} discussed the interworking issue of DSRC and cellular communication solutions. Perales \textit{et al.} \cite{Perales8642796} proposed a heterogeneous DSRC and mmWave vehicular network, which used side information of DSRC channel to speed up mmWave beam alignment. Prior work of Sim \textit{et al.} \cite{Sim8472783} integrated mmWave into C-V2X. Katsaros \textit{et al.} \cite{Katsaros7277110} developed stochastic network calculus to derive the stochastic upper bound of the end-to-end delay for the DSRC combining with C-V2X hybrid communications. 


However, to our best knowledge, there is no literature discussed the interworking of DSRC, C-V2X, and mmWave communications in the heterogeneous VEC system. In addition, existing work on VEC devoted to the computation offloading between the vehicle and the VEC servers. They ignored the offloading failure probability due to communication issues and the reliability of VEC. To fill the research gap, our work attempts to addresses the problem of tasks offloading in the heterogeneous VEC environment in accounts of the offloading failure probability and resource cost. This is a challenging problem due to the complicated dependency between multiple access technologies and offloading targets.

\section{System Model}

\begin{figure} [t]
     \centering
     \includegraphics[width=0.48\textwidth]{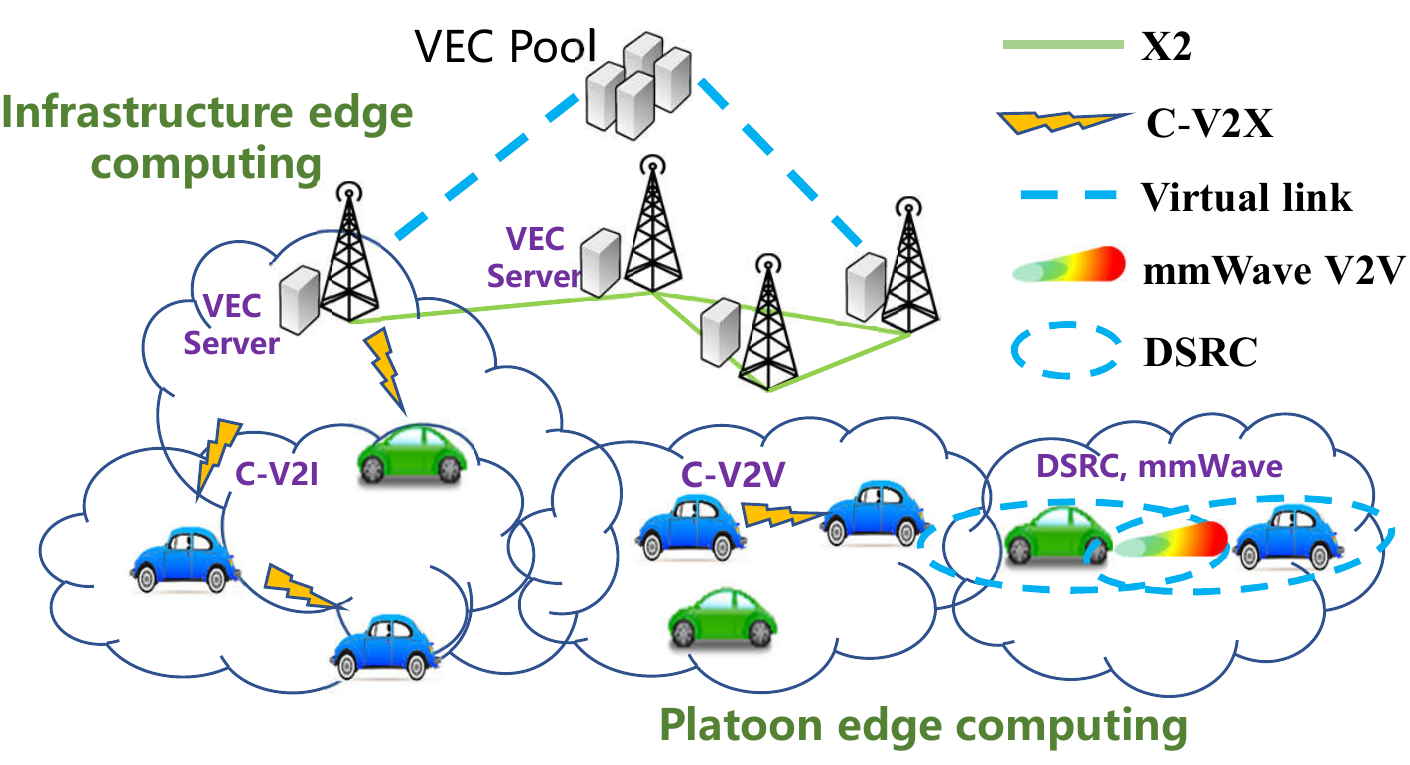} 
     \caption{Illustration of the heterogeneous VEC framework.} 
     \label{dsfjhjklhalllokoko}
\end{figure}

As shown in Fig.~\ref{dsfjhjklhalllokoko}, the proposed VEC framework consists of vehicles, cellular base stations (BSs), and VEC servers, where VEC servers are located with the BSs at the roadside to reduce the end-to-end communication and process latency. Since the C-V2I communication has several advantages over other V2I technologies, including longer range and enhanced reliability \cite{Sial8859331}, this paper only regards the cellular BS as the roadside infrastructure. In addition, VEC servers are connected with each other through the X2 interfaces of the accompanied BSs to form a resource pool, namely, the VEC pool. To simplify the model, we apply the virtual link to depict the logical connection between the VEC pool and the connected VEC servers. In this paper, the centralized resource management is implemented in the VEC pool that provides efficient resource utilization to the C-V2I offloading tasks. However, the resource management in a platoon is achieved by the platoon header that is a resource-rich vehicle elected from the platoon vehicles \cite{Qiao8879573}. 

\begin{figure*}
\centering
     \includegraphics[width=.9\textwidth]{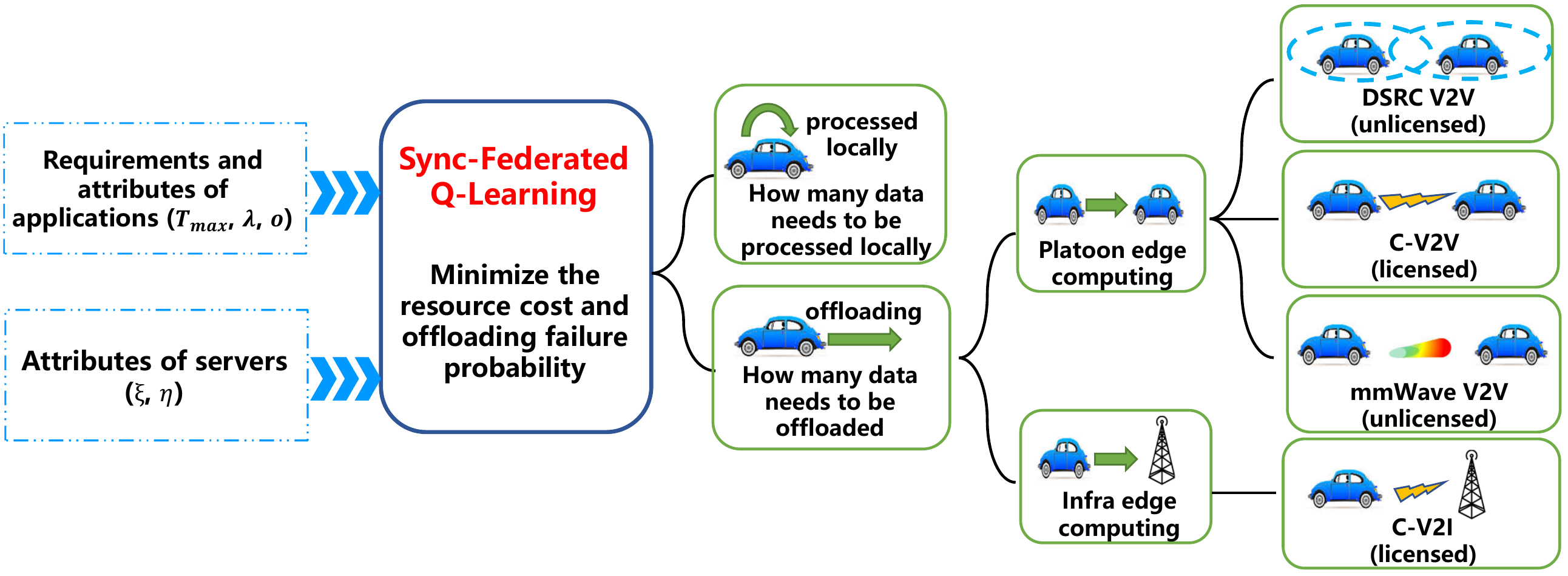} 
     \caption{Demonstration of the offloading scheduling. } 
\label{algAa}
\end{figure*}

Furthermore, we propose the Synchronous Federated Q-Learning (Sync-FQL) algorithm that instructs the offloading direction of the network traffic taking accounts of offloading failure probability and resource cost. This algorithm is deployed in the VEC pool or platoon header. As shown in Fig.~\ref{algAa}, the input of Sync-FQL algorithm includes delay requirements $T_{max}$ of the offloading tasks, traffic attributes of the offloading tasks (arrival rate $\lambda$ of and brustiness measure $o$), as well as the service attributes of servers (envelope service rate $\xi$ and peak service capacity $\eta$ of each server) \cite{Rizk6868978}. Note that we treat V2X communication and task processing as a service.

Since the V2X technologies and the offloading targets (vehicles or VEC servers) are tightly coupled, it is a typical studied object for the stochastic network calculus theory. Recent advances in network performance researches have adopted network calculus to estimate the end-to-end delay \cite{{Katsaros7277110},{Yang8252754}}. However, these previous works did not investigate the stochastic network calculus in the VEC system from the perspective of the offloading failure probability as well as communication and computing resources. In this paper, we investigate the stochastic network calculus to obtain the offloading performance metrics. Once the theoretical performance of each V2X technology is available, it can be used to guide the search pruning in the training process. 


Moreover, offloading scheduling is impacted by the properties of input network traffic. Thus, it is important to capture the statistical characteristics of the input traffic. From the perspective of the network calculus, an arrival traffic curve $X(t)$ can be modelled as the cumulative volume of the input traffic during interval $(0,t]$. And, $X(\tau,t) = X(t- \tau) = X(t) - X(\tau)$ is the cumulative volume of the input traffic during $(\tau,t]$, where $\tau<t$. Moreover, $X(\tau,t)$ has a statistical envelope $(\lambda, o)$, referred to as exponentially bounded burstiness model \cite{Rizk6868978}, which provides a validated inequality $P[X_{i}(t-\tau) \ge \lambda_i\cdot (t - \tau)+o_i] \le \varepsilon$, where $\lambda_i$ is an arrival rate of task $i$. $o_i$ is the burstiness measure of task $i$ \cite{Rizk6868978}. And, $\varepsilon$ is the violated probability. In virtue of Chernoff's bound \cite{Rizk6868978}, we get
\begin{equation}
\begin{split}
P[X_{i}(t-\tau) \ge \lambda_i\cdot (t - \tau)+o_i] \le e^{-\theta (\lambda_i\cdot (t - \tau)+o_i)} \mathrm{E}\left[e^{\theta X_{i}}\right],
\end{split}
\label{assafsafdsgesrewrb}
\end{equation}

\noindent where $\mathrm{E}(x)$ is the expectation of random variable $x$. $\theta$ is a constant parameter. We assume that there are $K$ categories automotive tasks $\{1,\dots,K\}$ in the transportation system, such as vehicular Internet and infotainment, remote diagnostics and management, cooperative lane change assist, and cooperative adaptive cruise control \cite{Campolo2017}. Each task $i$ generates own network traffic that is characterized by the arrival rate $\lambda_i$ and brustiness measure $o_i$. In addition, the task can be divided into several parts and processed separately.

Hereafter, we investigate the dynamic service curve that is provided by a channel or a processor $Y(\tau, t), \  t \ge \tau \ge 0$ \cite{Jiang2008Stochastic}. $Y(\tau, t)$ is non-negative and increases with $t$. A dynamic service envelope is defined as $P[Y(t-\tau) \ge \xi\cdot (t - \tau) - \eta] \le \varepsilon$ for all $t\ge\tau\ge 0$, where $\xi$ is the envelope service rate. $\eta$ is the peak service capacity of the server \cite{Rizk6868978}. According to Chernoff's bound, the envelope is rewritten as  
\begin{equation}
\begin{split}
P[Y(t-\tau) \le \xi\cdot (t - \tau) - \eta] \le e^{\theta (\xi\cdot (t - \tau) - \eta)} \mathrm{E}\left[e^{-\theta Y}\right].
\end{split}
\label{skladjl}
\end{equation}  

\noindent When the dynamic server $Y(s, t)$ processes the input traffic $X(s, t)$, the processing delay $d(\varepsilon)$ with a failure probability $\varepsilon(\eta)$ satisfies the following inequality \cite{Fidler4015760} 
\begin{equation}
\begin{split}
\begin{aligned}
\mathbb{P}\left[\left(X \oslash Y_{\mathrm{net}}\right)(t+d(\varepsilon), t) \geq 0\right] \leq \varepsilon(\eta),
\end{aligned}
\end{split}
\label{sdgjhbaeslkfjl}
\end{equation} 

\noindent where $X \oslash Y(s,t) = \sup _{\tau \in[0, s]}[X(\tau, t)-Y(\tau, s)]$.


\section{Performance Analysis on VEC}

In this section, we derive the end-to-end delay upper bound of different networks in the platoon-based edge computing and infrastructure-based edge computing, respectively.

\subsection{Platoon-based Edge Computing}

Vehicles can form a platoon to share their computing resource with the surrounding vehicles for cooperatively processing tasks. The offloading target of the platoon-based edge computing is the neighbor vehicles. There are three kinds of communication technologies to support the platoon-based edge computing that is shown in Fig.~\ref{algAa}. The C-V2V communication operates in a licensed band and needs the vehicles to pay a fee to the mobile network operator. While DSRC and mmWave V2V communications are free to access.


\subsubsection{\textbf{Delay Upper Bound of DSRC}}

The DSRC standard is based on the 802.11p amendment to the IEEE 802.11 standard that adapts the exponential back-off algorithm to cope the access competition. The initial size of the back-off window is assumed as $W_0$. The retry limit is set to $\gamma$. Thus, the number of the back-off stage is $\gamma + 1$ \cite{Katsaros7277110}. And, the available bandwidth of DSRC is denoted by $R^{dsrc}$. Therefore, the size of the window in the back-off state ${g}$ can be expressed as $W_\mathsf{g}=\min\left\{   2^{{{g}}} W_0,2^{\mathscr{G}}W_0\right\}$, where $\mathscr{G}$ is a threshold of the back-off counter, $0 < \mathscr{G} \le \gamma$. When the back-off counter exceeds $\mathscr{G}$, the size of the window will not grow anymore. Hence, the access delay of input traffic at the head-of-line is $\hat{t}_{\mathrm{serv}}=\sum_{g=0}^{\gamma} \hat{t}_{g}$, in which $\hat{t}_g$ is the longest duration of the back-off stage $g$. And, $\hat{t}_{g}=w^{B}_g t_w+t_{T X}$ \cite{Dianati7277110}, where
\begin{equation}
\begin{split}
\begin{aligned}
{w}^B_{g}=\left\{\begin{array}{ll}{2^{g}W_0} & g\le\mathscr{G} \\ {2^{\mathscr{G}}W_0} & {g>\mathscr{G}} \end{array}\right.,
\end{aligned}
\end{split}
\label{mnaesfiuk}
\end{equation} 

\noindent represents the largest number of back-off slots at stage $g$. $t_w$ is the average length of a back-off slot. And, $t_{TX} = \frac{{o}}{R^{dsrc}}$ is the average length of a transmission slot, where ${o}$ is the average brustiness measure of the input traffic \cite{Rizk6868978}. Without loss of generality, we assume $t_w = 1$ and the probability of a collision occurring in a transmission slot is $\mathscr{M}$. Therefore, $\frac{\mathscr{M}}{R^{dsrc}}$ represents the probability of a collision occurring in an unit period. If $\frac{2\mathscr{M}}{R^{dsrc}} > 1$, we get $\hat{t}_{\mathrm{serv}} \approx  W_0(\frac{2\mathscr{M}}{R^{dsrc}})^{\mathscr{G}}$. 

Moreover, the DSRC transmission process can be regarded as the classical latency-rate service $\beta(t)=R(t-s)^+$ where $s$ is the access delay \cite{Jiang2008Stochastic}. Therefore, the latency-rate service curve of DSRC is given as
\begin{equation}
\begin{split}
\begin{aligned}
\beta^{dsrc}(t-\tau) =R^{dsrc} \left((t-\tau)-\hat{t}_{\mathrm{serv}}\right)^+ ,
\end{aligned}
\end{split}
\label{dakhkj}
\end{equation} 

\noindent in which $(X)^+=0$ when $X<0$. Otherwise, $(X)^+ = X$. By virtue of the superposition property \cite{Rizk6868978}, the whole input traffic except for task $i$ is regarded as the background arrival curve $\alpha^{bg}_{i}(t)$ for task $i$:
\begin{equation}
\begin{split}
\begin{aligned}
{\alpha}^{bg}(t-\tau)=\sum\limits_{j\neq i}^{K}\varrho^{dsrc}_j \mathscr{O}_j(t-\tau),
\end{aligned}
\end{split}
\label{ioroipqw}
\end{equation}

\noindent where $\varrho^{dsrc}_j$ is the percentage of task $j$ using the DSRC offloading. And, $\mathscr{O}_j(t-\tau)$ is the accumulated traffic volume of task $j$ in interval $[\tau, t]$, i.e., $\mathscr{O}_j(t-\tau) =\lambda_j\cdot (t - \tau)+o_j$, where $\lambda_j$ is an envelope arrival rate of task $j$; $o_j$ is a burstiness measure of task $j$ \cite{Rizk6868978}. Then, based on the theory of Leftover Service \cite{Jiang2008Stochastic}, the service curve $S^{dsrc}_i(t-\tau)$ of DSRC transmission for task $i$ is expressed as
\begin{equation}
\begin{split}
\begin{aligned}
& S_i^{dsrc}(t-\tau) = (\beta^{dsrc} - \alpha^{bg}_i)^+ (t - \tau)
\\ &= (R^{dsrc}-\sum\limits_{j\neq i}^K \varrho^{dsrc}_j\lambda_j) (t-\tau) - R^{dsrc}\hat{t}_{\mathrm{serv}} - \sum\limits_{j\neq i}^K\varrho^{dsrc}_j o_j  . 
\end{aligned}
\end{split}
\label{oooooooroio}
\end{equation} 

\noindent If task $i$ has offloaded to the target vehicle, it needs to compete with other offloaded tasks (i.e., offloaded by mmWave, DSRC, and C-V2V) and the local processed tasks for the on-board CPU cycling (computing resource). Similar to Eq.~(\ref{ioroipqw}), the background on-board processing service curve of task $i$ is
\begin{equation}
\begin{split}
\begin{aligned}
{\phi}^{bg}_{i}(t-\tau)&=\frac{\mathscr{O}_j(t-\tau) }{N}\left[\sum_{j=1}^{K}(1-\varrho^{v2i}_j)-\varrho^{dsrc}_i\right],
\end{aligned}
\end{split}
\label{pwpwpwppwpwpw}
\end{equation}


\noindent where $N$ is the number of vehicles in the road segment. $\varrho_j^{x}$ is the percentage of the task $j$ using $x$ category offloading, where $x$ represents one of the DSRC, C-V2I, C-V2V, mmWave, and local processing. Hence, for any task $i$, the equation $\varrho^{local}_i+\varrho^{mmw}_i+\varrho^{dsrc}_i+\varrho^{v2i}_i+\varrho^{cv2v}_i = 1$ holds. Afterwards, based on the Leftover Service property \cite{Jiang2008Stochastic}, the on-board processing service curve for task $i$ using DSRC offloading is given as
\begin{equation}
\begin{split}
\begin{aligned}
&\Omega^{dsrc}_{i}(t-\tau)=(\Omega^{veh}- {\phi}^{bg}_{i})^+(t-\tau)
\\&= \left[\Theta^{veh}-\frac{1}{N}(\sum\limits_{j\neq i}^{K}\varrho^{dsrc}_{j}\lambda_j+\sum\limits_{j=1}^{K} (1 -\varrho^{dsrc}_j +\varrho^{v2i}_j) \lambda_j)\right] \cdot 
\\&(t-\tau) -\frac{1}{N}\left(\sum\limits_{j\neq i}^{K}\varrho^{dsrc}_{j} o_j+ \sum\limits_{j=1}^{K} (1 -\varrho^{dsrc}_j +\varrho^{v2i}_j )o_j\right),
\end{aligned}
\end{split}
\label{ckdckmmmmmmdls}
\end{equation} 

\noindent where $\Omega^{veh}(t-\tau)=\Theta^{veh} \cdot (t-\tau)$, and $\Theta^{veh}$ is the computing capacity for the on-board processor. Furthermore, according to the concatenated property \cite{Fidler4015760}, the total service curve of the DSRC transmission and the on-board processing is 
\begin{equation}
\begin{split}
\begin{aligned}
Y_{i}^{dsrc}(t-\tau) = (S_i^{dsrc} \otimes \Omega^{dsrc}_{i})(t-\tau),
\end{aligned}
\end{split}
\label{asdashghkdkjdfgaewjhfgegfru}
\end{equation} 

\noindent which represents task $i$ traversing the services of DSRC transmission and the on-board processor, where $(x \otimes y)(s, t) =\inf _{\tau \in[s, t]}[x(s, \tau)+y(\tau, t)]$. Therefore, based on Eq.~(\ref{sdgjhbaeslkfjl}), the validated inequality of DSRC offloading is
\begin{equation}
\begin{split}
\begin{aligned}
&\mathbb{P} \{ X_i \oslash Y_i^{ {dsrc}}(t+d(\varepsilon), t) \ge 0\}  \\& \le e^{\theta \cdot 0} \mathrm{E} e^{\theta[\sup\limits_{\tau\in[0, t]} X_i(\tau, t)-Y^{dsrc}_i(\tau, t+d(\varepsilon))]}
\\ & \le e^{\theta \varrho^{dsrc}_i o_i}  \sum\limits_{\tau\in [0,t]} e^{\theta\lambda_i(t-\tau)}  \sum_{v=\tau}^{t} \mathrm{E}\left[e^{-\theta S^i_{dsrc}(\tau, v)}\right]\cdot
\\ & \mathrm{E}\left[e^{-\theta \Omega^{dsrc}_{i}(v, t+d(\varepsilon))}\right]
\\ & \le  \frac{e^{\theta \varrho^{dsrc}_i o_i+\theta (\eta^{comp}_{dsrc}+\eta^{dsrc})} e^{-\theta \xi^{comp}_{dsrc} d(\varepsilon)}}{1-e^{-\theta ((\xi^{dsrc}-\xi^{comp}_{dsrc}))}}  \sum\limits_{\tau=0}^{\infty}  e^{-\theta (\xi^{comp}_{dsrc}-\lambda_i)\tau}
\\ & =  \frac{e^{\theta \varrho^{dsrc}_i o_i+\theta (\eta^{comp}_{dsrc}+\eta^{dsrc})} e^{-\theta \xi^{comp}_{dsrc} d_i}}{(1-e^{-\theta ((\xi^{dsrc}-\xi^{comp}_{dsrc}))})(1-e^{-\theta (\xi^{comp}_{dsrc}-\lambda_i)})},
\end{aligned}
\end{split}
\label{bllllbbbbbbbbbbbbl}
\end{equation}

\noindent where $\xi^{dsrc} = R^{dsrc} $, $\eta^{dsrc} =R^{dsrc}\hat{t}_{\mathrm{serv}}$, $\xi^{comp}_{dsrc} = \Theta-\frac{1}{N}[\sum\limits_{j\neq i}^{K}\varrho^{dsrc}_{j}\lambda_j+\sum\limits_{j=1}^{K} (\varrho^{mmw}_j +\varrho^{cv2v}_j +\varrho^{local}_j) \lambda_j] $, and $\eta^{comp}_{dsrc} = \frac{1}{N}[\sum\limits_{j\neq i}^{K}\varrho^{dsrc}_{j} o_j+ \sum\limits_{j=1}^{K} (\varrho^{mmw}_j + \varrho^{cv2v}_j +\varrho^{local}_j )o_j]$. The first inequality of Eq.~(\ref{bllllbbbbbbbbbbbbl}) is attained by the Chernoff's bound. And, the second inequality is derived by the fact:
\begin{equation}
\begin{split}
\begin{aligned}
\mathrm{E}\left[e^{-\theta Y_i^{dsrc}(\tau, t)}\right]&=\mathrm{E}\left[e^{-\theta \min _{v \in[\tau, t]}\left\{S_i^{dsrc}(\tau, v)+\Omega^{dsrc}_{i}(v, t)\right\}}\right] \\ & \leq \sum_{v=\tau}^{t} \mathrm{E}\left[e^{-\theta S_i^{dsrc}(\tau, v)}\right] \mathrm{E}\left[e^{-\theta \Omega^{dsrc}_{i}(v, t)}\right].
\end{aligned}
\end{split}
\label{AKLJDklioew_n}
\end{equation} 
 
\noindent The third inequality of Eq.~(\ref{bllllbbbbbbbbbbbbl}) is based on the assumption $\mathrm{E}\left[e^{-\theta Y(\tau, t)}\right] \leq e^{-\theta(\xi(t-\tau)-\eta)}$ with $t \rightarrow \infty$. The last equality of Eq.~(\ref{bllllbbbbbbbbbbbbl}) is obtained by the infinity geometric sum, where $(\xi^{dsrc}-\xi^{comp}_{dsrc})>0$ is assumed for convergence. In addition, we can regard the last equation as the offloading failure probability of task $i$:
\begin{equation}
\begin{split}
\begin{aligned}
\varepsilon_i = \frac{e^{\theta \varrho^{dsrc}_i o_i+\theta (\eta^{comp}_{dsrc}+\eta^{dsrc})} e^{-\theta \xi^{comp}_{dsrc}d_i}}{(1-e^{-\theta (\xi^{dsrc}-\xi^{comp}_{dsrc})})(1-e^{-\theta (\xi^{comp}_{dsrc}-\lambda_i)})},
\end{aligned}
\end{split}
\label{yyyyyuyyyyyu}
\end{equation}

\noindent where $\varepsilon_i$ represents the probability that the offloading delay (transmission delay plus the on-board processing delay) exceeds the value of $d_i$. Through a simple algebraic operation, we get the upper bound of the offloading delay $d_i$:
\begin{equation}
\begin{split}
\begin{aligned}
d_i &=\frac{1}{\xi^{comp}_{dsrc} } (\varrho_i^{dsrc} o_i+\eta^{comp}_{dsrc}+\eta^{dsrc}) -\frac{1}{\theta\xi^{comp}_{dsrc} } \left(\ln \varepsilon  + \mathscr{U}\right)
\\ & \le \frac{1}{\xi^{comp}_{dsrc} } (\varrho_i^{dsrc} o_i+\eta^{comp}_{dsrc}+\eta^{dsrc}-\frac{\ln\varepsilon}{\theta}),
\end{aligned}
\end{split}
\label{sdahoireawfo}
\end{equation} 

\noindent where $\mathscr{U} = \ln (1-e^{-\theta ((\xi^{dsrc}-\xi^{comp}_{dsrc}))}) + \ln (1-e^{-\theta (\xi^{comp}_{dsrc}-\lambda_i)})$. The inequality of Eq.~(\ref{sdahoireawfo}) is because of $\mathscr{U} \le 0$. Hence, the delay upper bound of DSRC offloading for task $i$ with the offloading failure probability $\varepsilon_i$ is expressed as
\begin{equation}
\begin{split}
\begin{aligned}
&d^{dsrc}_i =\\
& \frac{\frac{1}{N}\sum\limits_{j =1}^{K}[1-\varrho_j^{v2i}+(N-1)\varrho^{local}_j] o_j+\frac{W_0({2\mathscr{M}})^{\mathscr{G}}}{(R^{dsrc})^{\mathscr{G}-1}} -\frac{\ln\varepsilon_i}{\theta}}{\Theta^{veh}-\frac{1}{N}\sum\limits_{j =1}^{K}[1-\varrho_j^{v2i}+(N-1)\varrho^{local}_j]\lambda_j + \varrho^{dsrc}_i\lambda_i }.
\end{aligned}
\end{split}
\label{wkwkwkwkwwww}
\end{equation}

\subsubsection{\textbf{Delay Upper Bound of C-V2V}}

C-V2V communication can be regarded as specific reservation-based V2V technology. Additionally, Release 14 of C-V2X standardization in the Third Generation Partnership Project defines two new modes (mode 3 and mode 4) for C-V2V communication \cite{Chen7992934}. To simplify the analysis, we focus on the C-V2V communication under mode 3. In mode 3, the communication resource for each V2V communication is pre-assigned by the nearby base station. There is no access competition in the transmission procedure. Hence, the service curve of C-V2V communication under mode 3 is $\beta_i^{{cv2v}} = R^{cv2v}_i  (t-\tau)$, where $R^{cv2v}_i$ is the exclusive bandwidth for task $i$.  Similar to the DSRC task offloading, task $i$ also needs to compete with other offloaded tasks and the local processed tasks in the on-board processor. Based on the Leftover Service property, the on-board processing service curve of C-V2V offloading is  
\begin{equation}
\begin{split}
\begin{aligned}
&\Omega^{cv2v}_{{i}}(t-\tau) =
\\&  [\Theta^{veh} + \varrho^{cv2v}_i \lambda_i  - \frac{1}{N}\sum\limits_{j=1}^K(1 - \varrho^{v2i}_j + (N-1) \varrho^{local}_j )\lambda_j ] \cdot 
\\ &(t - \tau)  - \left\{  \frac{1}{N}\sum\limits_{j=1}^K (1- \varrho^{v2i}_j + (N-1) \varrho^{local}_j) o_j - \varrho^{cv2v}_i o_i\right\}.
\end{aligned}
\end{split}
\label{la_fa_ss}
\end{equation}

\noindent Hereafter, by virtue of the concatenated property, the total service curve of C-V2V offloading for task $i$ is $Y^{cv2v}_{{i}}(t-\tau)  = \beta^{cv2v}_i \otimes \Omega^{cv2v}_{{i}} (t-\tau)$. Therefore, the validated inequality of C-V2V offloading is given as
\begin{equation}
\begin{split}
\begin{aligned}
&\mathbb{P} \{ X_i \oslash Y^{cv2v}_i(t+d_i, t) \ge 0\} 
\\& \le  \frac{e^{\theta \varrho^{cv2v}_i o_i + \theta (\eta^{comp}_{cv2v}+\eta^{cv2v})} e^{-\theta \xi^{comp}_{cv2v} d(\varepsilon)}}{(1-e^{-\theta ((\xi^{cv2v}-\xi^{comp}_{cv2v}))})(1-e^{-\theta (\xi^{comp}_{cv2v}-\lambda_i)})} .
\end{aligned}
\end{split}
\label{farlarroioiooioi}
\end{equation} 

\noindent where $\xi^{cv2v} = R_i^{cv2v}$, $\xi^{comp}_{cv2v} = \Theta^{veh} + \varrho^{cv2v}_i \lambda_i  - \frac{1}{N}\sum\limits_{j=1}^K(1 - \varrho^{v2i}_j + (N-1) \varrho^{local}_j )\lambda_j $, $\eta^{comp}_{cv2v} = -\varrho^{cv2v}_i o_i  + \frac{1}{N}\sum\limits_{j=1}^K (1- \varrho^{v2i}_j + (N-1) \varrho^{local}_j) o_j$, and $\eta^{cv2v} = 0$. Then, we denote the right hand-side of Eq.~(\ref{farlarroioiooioi}) as the offloading failure probability $\varepsilon_i$. Hence, the delay upper bound $d^{cv2v}_i$ of C-V2V offloading is
\begin{equation}
\begin{split}
\begin{aligned}
&d^{cv2v}_i =  \\
& \frac{\frac{1}{N}\sum\limits_{j =1}^{K}(1-\varrho_j^{v2i}+(N-1)\varrho_j^{local})o_j-\frac{\ln\varepsilon_i}{\theta}}{\Theta^{veh} - \frac{1}{N}\sum\limits_{j =1}^{K}(1-\varrho_j^{v2i}+(N-1)\varrho^{local}_j)\lambda_j + \varrho^{cv2v}_i\lambda_i} ,
\end{aligned}
\end{split}
\label{larrxdddddddddi}
\end{equation}

\subsubsection{\textbf{Delay Upper Bound of mmWave}}
The implementation of mmWave communication needs the antenna beam alignment at first that incurs additional time overhead compared to other V2V communications \cite{chongwenTVT}. Furthermore, the usage of the control channel for beam alignment can significantly reduce the complexity of the alignment \cite{Gruteser8642796}. This control channel delivers the Request-To-Send like (RTS-like) and Clear-To-Send like (CTS-like) beacons that contain vehicular kinetic information (location, speed, acceleration, and heading direction) and the communication types \cite{Gruteser8642796}. In addition, the sub-6GHz control channel for RTS/CTS-like beacons can be a competition-based channel or a reservation-based channel. To simplify the notations, the mmWave offloading with the competition-based channel aided alignment is denoted by CmmW. And, RmmW represents the mmWave offloading with the reservation-based channel aided alignment.

\paragraph{\textbf{Delay Upper Bound of CmmW}} 

\begin{figure} [t]
     \centering
     \includegraphics[width=0.48\textwidth]{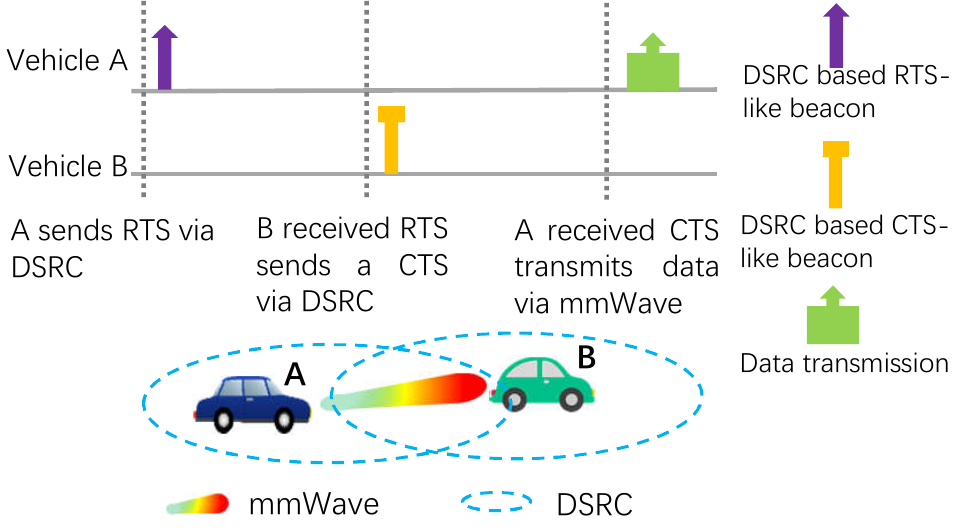} 
     \caption{Beam alignment of CmmW communications. } 
     \label{osacascojaooooo}
\end{figure}

In this sub-section, we resort to the DSRC channel as the competition-based control channel for the beam alignment, which is depicted in Fig.~\ref{osacascojaooooo}. The mmWave communication includes two procedures: the beam alignment and the data transmission. Vehicle $A$ initiates the beam alignment through sending an RTS-like beacon that contains the kinetic information of vehicle $A$. When vehicle $B$ has listened to the RTS-like beacon, it returns a CTS-like beacon with its kinetic information to confirm the mmWave communication, then the alignment procedure has completed \cite{Perales8642796}. Due to the concatenated property, the service curve of mmWave communication is the convolution of the beam alignment procedure and the data transmission procedure. Moreover, the beam alignment procedure is composed of the RTS-like beacon transmission and CTS-like beacon transmission. Consequently, the service curve of mmWave communication is given as
\begin{equation}
\begin{split}
\begin{aligned}
{\beta^{{net}}(t-\tau)=\left(\beta^{{RTS}} \otimes \beta^{ {CTS}} \otimes \beta^{{CmmW }}\right)(t - \tau)},
\end{aligned}
\end{split}
\label{iiiiieeeeeeeeei}
\end{equation}

\noindent where $\beta^{{RTS}}$ and $\beta^{{CTS}}$ are the service curves of RTS and CTS transmissions, respectively. In general, the transmissions of RTS and CTS are symmetrical. Thus, we assume the service curves of RTS and CTS transmissions are same. However, due to the beam alignment using the DSRC channel, the original DSRC traffic could impact on the alignment performance. According to the Leftover Service theory, the service curve of the DSRC-based RTS is $\beta^{ {RTS}} =[R^{dsrc}_i -  (2\mathscr{O}^{b}+\sum\limits_{j=1}^N\varrho^{dsrc}_j\mathscr{O}_j)]^+(t-\tau)$, where $\mathscr{O}^{b}$ is the brustiness measure of the RTS-like traffic that is identical to that of the CTS-like traffic. However, the brustiness measure of RTS/CTS-like beacons usually too small to interfere with the channel capacity, compared with that of the DSRC traffic. To simplify the analysis, the brustiness measure of RTS/CTS-like traffic is negligible, i.e., $\mathscr{O}^{b} = 0$. In addition, the service curve of mmWave data transmission is $\beta^{CmmW}(t - \tau) = R^{mmW} \times(t-\tau)$, where $R^{mmW}$ is the channel capacity of mmWave. Therefore, according to Eq.(\ref{iiiiieeeeeeeeei}), the service curve for mmWave transmission is
\begin{equation}
\begin{split}
\begin{aligned}
\beta^{net}(t-\tau) =[R^{dsrc}  -  2 \sum\limits_{j=1}^N\varrho^{dsrc}_j\mathscr{O}_j]^+(t-\tau).
\end{aligned}
\end{split}
\label{osososoosososo}
\end{equation} 

\noindent When task $i$ has arrived at the service vehicle through mmWave transmission, it also competes with other offloaded or local processed tasks for the on-board CPU cycling. The on-board processing service curve of the CmmW offloading for task $i$ is 
\begin{equation}
\begin{split}
\begin{aligned}
&\Omega^{CmmW}_{{i}}(t-\tau) =
\\& [\Theta^{veh} + \varrho^{CmmW}_i \lambda_i  - \frac{1}{N}\sum\limits_{j=1}^K(1 - \varrho^{v2i}_j + (N-1) \varrho^{local}_j )\lambda_j ] \cdot 
\\ &(t - \tau) -\left\{\frac{1}{N}\sum\limits_{j=1}^K (1- \varrho^{v2i}_j + (N-1) \varrho^{local}_j) o_j - \varrho^{CmmW}_i o_i \right\}.
\end{aligned}
\end{split}
\label{koqlaicakpewjdc}
\end{equation} 

\noindent Hence, based on the concatenated property, we get the total service curve of the CmmW offloading, $Y^{CmmW}_i$:
\begin{equation}
\begin{split}
\begin{aligned}
Y^{CmmW}_i(t-\tau) = \beta^{net}\otimes \Omega^{CmmW}_{{i}}(t-\tau).
\end{aligned}
\end{split}
\label{lklklklkl}
\end{equation}  

\noindent The validated inequality for the CmmW offloading is 
\begin{equation}
\begin{split}
\begin{aligned}
&\mathbb{P} \{ X_i \oslash Y^{CmmW}_i(t+d(\varepsilon), t) \ge 0\} 
\\& \le  \frac{e^{\theta \varrho^{mmw}_i o_i + \theta (\eta^{comp}_{mmw}+\eta^{mmw})} e^{-\theta \xi^{comp}_{mmw} d(\varepsilon)}}{(1-e^{-\theta ((\xi^{mmw}-\xi^{comp}_{mmw}))})(1-e^{-\theta (\xi^{comp}_{mmw}-\lambda_i)})}  ,
\end{aligned}
\end{split}
\label{wwwwwwwmsflkarlk}
\end{equation} 

\noindent where $\xi^{mmw} = R^{dsrc}  -  2 \sum\limits_{j=1}^N\varrho^{dsrc}_j\mathscr{O}_j$, $\xi^{comp}_{mmw} = \Theta^{veh} + \varrho^{mmw}_i \lambda_i  - \frac{1}{N}\sum\limits_{j=1}^K(1 - \varrho^{v2i}_j + (N-1) \varrho^{local}_j )\lambda_j $, $\eta^{comp}_{mmw} = -\varrho^{mmw}_i o_i  + \frac{1}{N}\sum\limits_{j=1}^K (1- \varrho^{v2i}_j + (N-1) \varrho^{local}_j) o_j$, and $\eta^{mmw} = 2 \sum\limits_{j =1}^{K} \varrho^{dsrc}_j o_j$. Since the right hand-side of Eq.~(\ref{wwwwwwwmsflkarlk}) can be regarded as the failure probability $\varepsilon_i$, the upper bound of CmmW offloading delay $d^{mmw_c}_i$ for task $i$ is
\begin{equation}
\begin{split}
\begin{aligned}
&d^{mmw_c}_i =
\\& \frac{2 \sum\limits_{j =1}^{K} \varrho^{dsrc}_j o_j+\frac{1}{N}\sum\limits_{j =1}^{K} (1-\varrho^{v2i}_j+(N-1)\varrho^{local}_j) o_j - \frac{\ln\varepsilon_i}{\theta}}{\Theta^{veh} - \frac{1}{N}\sum\limits_{j =1}^{K} (1-\varrho_j^{v2i}+(N-1)\varrho^{local}_j)\lambda_j+\varrho^{mmw}_i\lambda_i}.
\end{aligned}
\end{split}
\label{OkjlKKLHJHnm}
\end{equation} 

Compared with the C-V2V offloading, the mmWave offloading has an additional item $2 \sum\limits_{j =1}^{K} \varrho^{dsrc}_j o_j$ that causes by the DSRC traffic. The additional DSRC traffic deteriorates the performance of CmmW offloading.

\paragraph{\textbf{Delay Upper Bound of RmmW}} 

The reservation-based control channel is purchased from the mobile network operator or licensed by the standards in the future. Following the same derivation of the CmmW offloading, the service curve of RTS/CTS transmission in the RmmW alignment is $\beta^{ {RTS}}_r =[R^{rc}_i - 2\sum\limits_{j=1}^N\varrho^{mmw}_j\mathscr{O}^{b}_j ]^+ (t-\tau)$ where $R^{rc}$ is the capacity of reserved control channel. Therefore, the service curve $\beta^{net}_r$ for the RmmW transmission is
\begin{equation}
\begin{split}
\begin{aligned}
\beta^{net}_r(t-\tau) & =[R^{rc}  -  4 \sum\limits_{j=1}^N\varrho^{mmw}_j\mathscr{O}^{b}_j]^+(t-\tau).
\end{aligned}
\end{split}
\label{hjgdagdhagkesagkjgesuy}
\end{equation} 

\noindent In addition, the on-board processing service curve of the RmmW offloading is the same as that of CmmW offloading. Thus, the delay upper bound of RmmW offloading $d^{mmw_r}_i$ for task $i$ is
\begin{equation}
\begin{split}
\begin{aligned}
&d^{mmw_r}_i =
\\& \frac{4 \sum\limits_{j=1}^N\varrho^{mmw}_j o^{b}_j +\frac{1}{N}\sum\limits_{j =1}^{K} (1-\varrho^{v2i}_j+(N-1)\varrho^{local}_j) o_j - \frac{\ln\varepsilon_i}{\theta}}{\Theta^{veh} - \frac{1}{N}\sum\limits_{j =1}^{K} (1-\varrho_j^{v2i}+(N-1)\varrho^{local}_j)\lambda_j+\varrho^{mmw}_i\lambda_i},
\end{aligned}
\end{split}
\label{fadkg;jrhgliqghr}
\end{equation}

\noindent where $o^b_j$ is the brustiness measure of the RTS/CTS-like traffic. Usually, the volume of the control traffic is constrained to be very small to mitigate the communication budgets. Hence, the constant $4 \sum\limits_{j=1}^N\varrho^{mmw}_j o^{b}_j$ is negligible. Hereafter, the delay upper bound of RmmW offloading is similar to that of the C-V2V offloading, i.e., Eq.(\ref{larrxdddddddddi}). Consequently, the performance of C-V2V offloading can be a good reference for that of RmmW. Therefore, we do not individually explore the RmmW offloading in the remainder of this paper.

\subsection{Infrastructure-based Edge Computing}

As shown in Fig.~\ref{dsfjhjklhalllokoko}, cellular base stations, deployed in the roadsides, collect data from vehicles, and deliver the data to the VEC pool for processing. Thus, the C-V2I offloading has two components: uplink transmission and VEC pool processing. Since C-V2I uplink is typical reservation-based communication, its service curve can be expressed as
\begin{equation}
\begin{split}
\begin{aligned}
\beta_i^{\text {uplink }}(t-\tau) = R^{v2i}_i (t-\tau),
\end{aligned}
\end{split}
\label{rereawfo}
\end{equation} 




\noindent where $R^{v2i}_i$ is the assignment uplink bandwidth for task $i$. As for the VEC pool processing, the computing resources of VEC pool are shared to all the upload tasks. Thus, the VEC pool processing service curve for task $i$ is 
\begin{equation}
\begin{split}
\begin{aligned}
\Omega^{ {v2i}}_i(t-\tau)= (\Theta^{epc} - \sum\limits_{j\neq i}^{K}\varrho^{v2i}_j\lambda_j)(t-\tau) - \sum\limits_{j\neq i}^{K}\varrho^{v2i}_jo_j,
\end{aligned}
\end{split}
\label{12321jhgsaiua}
\end{equation} 

\noindent in which $\Theta^{epc}$ is the total computing capacity of the VEC pool that is larger than the computing capacity of on-board processor $\Theta^{veh}$.
The total service curve of C-V2I offloading for task $i$ is denoted by $Y^{v2i}_i(t-\tau) = \beta_{i}^{\text {uplink }} \otimes  \Omega^{ {v2i}}_i(t-\tau)$. Referring to the derivation of Eq.~(\ref{bllllbbbbbbbbbbbbl}), the validated inequality of C-V2I offloading is
\begin{equation}
\begin{split}
\begin{aligned}
&\mathbb{P} \{ X_i \oslash Y_i^{v2i}(t+d(\varepsilon), t) \ge 0\}  \\
\\ & \le  \frac{e^{\theta \o_i+\theta (\eta^{epc}+\eta^{uplink})} e^{-\theta \xi^{epc}d_i}}{(1-e^{-\theta ((\xi^{uplink}-\xi^{epc}))})(1-e^{-\theta (\xi^{epc}-\lambda_i)})},
\end{aligned}
\end{split}
\label{opopopa_lfs}
\end{equation} 


\noindent where $\xi^{epc} = \Theta^{epc} - \sum\limits_{j\neq i}^{K}\varrho^{v2i}_j\lambda_j$, $\xi^{uplink} = R^{v2i}_i$, $\eta^{epc} = \sum\limits_{j\neq i}^{K}\varrho^{v2i}_jo_j$, and $\eta^{uplink} = 0$. Therefore, the  delay upper bound of C-V2I offloading is:
\begin{equation}
\begin{split}
\begin{aligned}
d^{v2i}_i &=\frac{\sum\limits_{j = 1}^{K}\varrho^{v2i}_jo_j-\frac{\ln\varepsilon_i}{\theta}}{ \Theta^{epc} - \sum\limits_{j\neq i}\varrho^{v2i}_j\lambda_j }.
\end{aligned}
\end{split}
\label{kkkkkksafa_lfs}
\end{equation} 

\subsection{Local Processing}

Vehicular applications could be self-digested by the vehicle's own on-board processor. However, the local processed task $i$ still has to compete with other offloaded and local tasks. Hence, the local processing service curve of task $i$ is
\begin{equation}
\begin{split}
\begin{aligned}
&Y^{local}_{{i}}(t-\tau) = 
\\& [\Theta^{veh} - \frac{1}{N}\sum\limits_{j=1}^K(1 - \varrho^{v2i}_j + (N-1) \varrho^{local}_j )\lambda_j ] (t - \tau) 
\\ & - \frac{1}{N}\sum\limits_{j=1}^K (1- \varrho^{v2i}_j + (N-1) \varrho^{local}_j) o_j.
\end{aligned}
\end{split}
\label{qoqoiwwiiwiwiwiww}
\end{equation} 

\noindent And, the corresponding validated inequality is
\begin{equation}
\begin{split}
\begin{aligned}
&\mathbb{P} \{ X_i \oslash Y^{local}_i(t+d(\varepsilon), t) \ge 0\} 
\\& \le  \frac{e^{\theta \varrho^{local}_i o_i + \theta (\eta^{comp}_{local}+\eta^{local})} e^{-\theta \xi^{comp}_{local} d(\varepsilon)}}{(1-e^{-\theta ((\xi^{local}-\xi^{comp}_{local}))})(1-e^{-\theta (\xi^{comp}_{local}-\lambda_i)})}  ,
\end{aligned}
\end{split}
\label{qQqqqqkajsgooooorlk}
\end{equation} 

\noindent where $\xi^{local} = 0$, $\xi^{comp}_{local} = \Theta^{veh}  - \frac{1}{N}\sum\limits_{j=1}^K(1 - \varrho^{v2i}_j + (N-1) \varrho^{local}_j )\lambda_j $, $\eta^{comp}_{local} =  \frac{1}{N}\sum\limits_{j=1}^K (1- \varrho^{v2i}_j + (N-1) \varrho^{local}_j) o_j$, and $\eta^{local} = 0$. Let the right hand-side of Eq.~(\ref{qQqqqqkajsgooooorlk}) equal to $\varepsilon_i$, the delay upper bound $d^{local}_i$ for local processing is given as
\begin{equation}
\begin{split}
\begin{aligned}
d^{local}_i =\frac{  \frac{1}{N}\sum\limits_{j =1}^{K} (1-\varrho^{v2i}_j+(N-1)\varrho^{local}_j) o_j - \frac{\ln\varepsilon_i}{\theta}}{\Theta^{veh} - \frac{1}{N}\sum\limits_{j =1}^{K} (1-\varrho_j^{v2i}+(N-1)\varrho^{local}_j)\lambda_j }.
\end{aligned}
\end{split}
\label{safdjlskfasdjkfmkwiowq}
\end{equation}

\section{Model Optimization}

This section proposes a new optimization model taking account of the considered communication and computing cost, as well as the failure probability. When vehicles have utilized C-V2X communication, the cellular operator could charge the fee $A_i$ per Mbps to vehicles for transmission service. Moreover, the communication cost $C^{comm}_i(t-\tau)$ of task $i$ is only generated by the C-V2X communication due to the licensed band, i.e., 
\begin{equation}
\begin{split}
\begin{aligned}
C^{comm}_i(t-\tau) =A_{i}( \varrho^{v2i}_i +  \varrho^{cv2v}_i )  \mathscr{O}_i(t-\tau).
\end{aligned}
\end{split}
\label{daiZXU9uiqew}
\end{equation} 

\noindent Regarding the computing cost, the unit computing costs of the VEC pool and on-board processor are $\mathscr{A}^{infra}_i$ per Mbps and $\mathscr{A}^{veh}_i$ per Mbps, respectively. The VEC pool computing cost is produced by C-V2I offloading, while the computing cost of the on-board processor is yielded by the DSRC offloading, the C-V2V offloading, and the mmWave offloading. Since the local processing only employs the local computing resource, it does not generate any cost of communication and computing. Thus, the computing cost for task $i$ is 
\begin{equation}
\begin{split}
\begin{aligned}
&C^{comp}_i(t - \tau) =
\\&   [\mathscr{A}^{infra}_{i} \varrho^{v2i}_i + \mathscr{A}^{veh}_{i} (\varrho^{cv2v}_i+\varrho^{mmw}_i+\varrho^{dsrc}_i)] \mathscr{H}_i \mathscr{O}_{i}(t-\tau),
\end{aligned}
\end{split}
\label{9939204230iu}
\end{equation} 
 
\noindent where $\mathscr{H}_i$ is the computation complexity of task $i$ \cite{Liu2018}. 

Furthermore, we originally treat the failure probability as another cost in the VEC because the failed offloading will tightly deteriorate the quality of service for automotive tasks. Moreover, the minimized cost model for the heterogeneous VEC is proposed in $P1$.

\begin{equation}
\begin{aligned}
\begin{split}
 \textbf{P1:} \qquad &\min_{\bm \varrho} \ {\sum\limits_{i=1}^K {C^{comm}_i(\bm \varrho)  + C^{comp}_i(\bm \varrho) + \delta_i \varepsilon_i }}\\
 \text{s.t.} ,\ \ \ \ \ 
&\text{C1:} \  \varrho^{h}_i, \varepsilon_i \in [0,1],  \  \ \sum\limits_{h}\varrho^{h}_i = 1, \ i \in \mathcal{I} \\
&\text{C2:} \ \sum\limits_{i=1}^{K}( \varrho^{v2i} +  \varrho^{cv2v} )\mathscr{O}_i(t)\le R^{cv2x} \\
&\text{C3:} \ \sum\limits_{i=1}^{K} \varrho^{dsrc}_i \mathscr{O}_i(t)\le R^{dsrc} \\
&\text{C4:} \  \bigcap\limits_{h} \left\{d^{h}_i \le T^{max}_i\right\}, \ i \in \mathcal{I} \\
\end{split} 
\end{aligned} 
\label{e5}
\end{equation}

\noindent where $h \in \{dsrc, v2i, cv2v, mmw, local\}$. $\bm \varrho  = [\varrho^{dsrc}, \varrho^{v2i}, \varrho^{cv2v}, \varrho^{mmw}, \varrho^{local}]$. $T^{max}_i$ is the delay requirement of task $i$. $d^h_i$ and the offloading failure probability $\varepsilon_i$ of different access technologies are attained by Eq.~(\ref{wkwkwkwkwwww}), (\ref{larrxdddddddddi}), (\ref{fadkg;jrhgliqghr}), (\ref{OkjlKKLHJHnm}), (\ref{kkkkkksafa_lfs}), (\ref{safdjlskfasdjkfmkwiowq}), respectively. $\bm\delta=[\delta_1,\delta_2,\dots \delta_K]$ represents the priority of application $i$. This paper first proposes the offloading failure probability as the optimization objective that is rational and neglected in previous literature. However, since $P1$ is non-convex that is difficult to solve.

According to Eq.~(\ref{wkwkwkwkwwww}), (\ref{larrxdddddddddi}), (\ref{OkjlKKLHJHnm}), (\ref{kkkkkksafa_lfs}), (\ref{safdjlskfasdjkfmkwiowq}), $\varepsilon_i$ is inversely proportional to $d^h_i$. When $\varepsilon_i$ has been minimized, constraint $C4$ becomes tight, i.e., $d^h_i = T^{max}_i$. Consequently, $C4$ can be removed by replacing $\ln\varepsilon_i$ as $\max\limits_{h} \ln \varepsilon^h_i(\bm \varrho)$. Thus, $P1$ transforms into $P2$, equivalently.
\begin{equation}
\begin{aligned}
\begin{split}
 \textbf{P2:} \qquad &\min_{\bm \varrho} \ {\sum\limits_{i=1}^K {C^{comm}_i(\bm \varrho)  + C^{comp}_i(\bm \varrho) + \delta_i \max\limits_{h} \ln\varepsilon^h_i(\bm \varrho)}}\\
 \text{s.t.} ,\ \ \ \ \ 
&\text{C1}, \text{C2}, \text{C3}.\\
\end{split} 
\end{aligned} 
\label{fakjhfkjahflffffffffffffffffffffffffffffffff}
\end{equation}


Considering the non-convex objective of $P2$, it is also difficult to directly solve. In this paper, we propose two canonical solutions to address $P2$. One is a parallel learning-based method, federated Q-learning (FQL) that is presented in Alg.~\ref{Ag1}, in which $N_s$ is the training times of the federated Q-learning. $CQ-table$ is the consensus Q-table for aggregating. The other canonical solution is Relaxation optimization that applies the relaxation trick to transfer the original non-convex problem $P2$ to a convex problem with low complexity.
\begin{algorithm} 
	\caption{Sync Federated Q-Learning (Sync-FQL)} \label{Ag1}
	Initialize action-state Q-table $Q_i$; CQ-table $CQ$; $\bm \varrho$\\
	\For{$j:$ $1$ to $N_s$} 
	{
	\For{each offload technology $i$ in parallel}
	{
		Selecting an action with max $Q_i$ or randomly rollout with a certain probability;\\
		Update reward $r_i$ using Eq.~(\ref{sackljafuiwerifepucnjhcbhjhvadjk}); compute local update of $Q_i$ using Eq.~(\ref{sdsddsfljqewl2312439});\\

	}
	
	\If{the aggregator receives all local updated $Q_i(t)$}
		{
				Update global $CQ(t+1)$ using Eq.~(\ref{asiuchusai9320320});

Set $Q_i(t+1) \leftarrow CQ(t+1)$ for all $i$;
		
		}		
	
	}

\end{algorithm}

%
%
%
%
%

\subsection{Federated Q-Learning}

The V2X offloading selection is modeled as a markov decision process, which consists of a tuple $\{T, A, P, R\}$ where $T$ and $A$ are the set of states and actions, respectively. Transition probability $P(t_i|t_{i-1}, a_{i-1})$ is the probability of a transition occurred where an agent enters state $t_i$ after taking action $a_{i-1}$ at state $t_{i-1}$. Reward $R(t_i, a_i)$ is a feedback of selecting action $a_{i}$ at state $t_i$ \cite{Hendrik2012}.

\subsubsection{\textbf{State}}
$x_t = \{\bm{\varrho}(t), W_t\}$ is a state of the offloading scheduling, where $\bm{\varrho}(t) = \{\varrho^{dsrc}_i, \varrho^{v2i}_i \dots , \varrho^{local}_i\}$ is the assigned proportion of different offloading technologies used in task $i$. $W_t = \{R^{cv2x}_r,R^{dsrc}_r\}$ is the reserved bandwidth for C-V2X and DSRC offloading. In addition, according to current assignment $\bm{\varrho}$, the available bandwidth for C-V2X and DSRC offloading are updated to $R_r^{cv2x} = R^{cv2x} - \sum\limits_{i=1}^{K}( \varrho^{v2i} +  \varrho^{cv2v} )\mathscr{O}_i(t)$, and $R^{dsrc}_r =R^{dsrc} - \sum\limits_{i=1}^{K} \varrho^{dsrc}_i \mathscr{O}_i(t)$, respectively.

\subsubsection{\textbf{Action}}
an action stands for a traffic assignment of different offloading technologies. When one kind of offloading technology increases/decreases $0.01$ (or $0.1$) percentage of traffic volume, the other offloading technologies will equally decrease/increase $\frac{0.01}{N^{total}_f - 1}$ ( or $\frac{0.1}{N^{total}_f - 1}$) percentage of traffic volume, respectively. $N^{total}_f$ is the total number of offloading technologies. The agent has a certain probability $p$ to select the action with the maximum value of the Q-Table. or randomly choose an action in the available action set with probability $1-p$.

\subsubsection{\textbf{Transition Probability}}
at each state, $P(t_i|{t_{i-1}}, a_{i-1})$ is non-zero except for the offloading percentage $\varrho^{h}_i$ of one offloading technology is equal to $1$ or $0$ that cannot increase or decrease, respectively. Otherwise, the current offloading percentage of the technology randomly adds one of the element in $\{0.1,0.01,-0.1,-0.01\}$ to update its offloading percentage.



\subsubsection{\textbf{Reward}}
When the proportion (offloading percentage) of offloading technology $i$ updates, the algorithm returns reward $r_i$ that represents the gain of selecting offloading technology $i$. Noted that the reward is a negative form of the total cost.

\begin{equation}
\begin{split}
\begin{aligned}
r_i  = {- \delta_i \max\limits_{h} \ln\varepsilon^h_i(\bm \varrho) - \left\{ \sum\limits_{i=1}^K {C^{comm}_i(\bm \varrho)  + C^{comp}_i(\bm \varrho)}\right\}     }.
\end{aligned}
\end{split}
\label{sackljafuiwerifepucnjhcbhjhvadjk}
\end{equation} 

\subsubsection{\textbf{Q-Table}}

Q-Table could be regarded as an action-state performance index function. In this paper, the Q-Table is a three-dimensional matrix whose size is $N^{total}_f \times K \times N_{\bm \varrho}$, where $N^{total}_f$ is the number of the offloading technology (including local processing). $K$ is the number of the application category. $N_{\bm \varrho}$ is the number of feasible $\varrho_i$, and the number of feasible $\varrho_i$ is determined by the accuracy of $\varrho_i$. In this paper, the accuracy is set to 0.01, which means $\varrho_i$ will increase/decrease with at least 0.01 in each iteration. The number of feasible $\varrho_i$ is 101, counting from 0 to 100. Therefore, the element $[i,j,k]$ of the Q-table represents the learning reward (according to Eq.~(\ref{sdsddsfljqewl2312439})) attained by the state that $k$ percentage of the $j^{th}$ category application is offloaded through the $i^{th}$ technology. Additionally, the Q-table is updated as follow

\begin{equation}
\begin{split}
\begin{aligned}
Q_{i}(x_t, a)&=\alpha \left\{r_{t} +\gamma \max _{a}\left[Q_{i}\left(x_{t-1}, a\right)\right]\right\}\\
&+(1-\alpha)Q_{i}(x_{t-1}, a),
\end{aligned}
\end{split}
\label{sdsddsfljqewl2312439}
\end{equation} 

\noindent where $\alpha$ is the learning rate. $\gamma$ is the discount factor. Each offloading technology $i$ has own local Q-Table $Q_i(t)$, where $t$ denotes the iteration time. At $t = 0$, the local Q-Tables of all technologies are initialized to the same value. When $t > 0$, the local Q-learning updates a new local Q-table $Q_i(t)$ based on the previous local Q-table $Q_i(t-1)$. The Q-learning for $i^{th}$ technology offloading optimization is referred to as the local learning $i$. After multiple local updates, a global aggregation is performed through the aggregator (deployed at the VEC server and the platoon head).

Moreover, we investigate two kinds of aggregations: the asynchronous aggregation and the synchronous aggregation. The synchronous aggregation operation that can implement the global aggregation when all local learnings have been updated. However, an asynchronous aggregation that executes the global aggregation when each local learning has been updated. After the global aggregation, the aggregator yields a consensus Q-table $CQ$ based on the local Q-tables:
\begin{equation}
\begin{split}
\begin{aligned}
&CQ(t) = \frac{(N^{update} - 1) \times CQ(t-1) +\frac{1}{N_{f}} \sum\limits_i^{N^{total}_f} Q_i(t)}{N^{update}},
\end{aligned}
\end{split}
\label{asiuchusai9320320}
\end{equation} 

\noindent where $Q_i(t)$ is the local Q-table for offloading technology $i$ at iteration $t$. $N^{update}$ is the number of the updated local Q-Table that is equal to $N^{total}_f$ in the synchronous aggregation. The synchronous update represents that when all local learnings have updated their own Q-table, the global consensus Q-table $CQ(t)$ is updated. In the simulation part, we can see that the performance of synchronous federated learning outperforms that of asynchronous federated learning. The main potential problem of the asynchronous aggregation is that each local update is achieved on the model that could be outdated. Consequently, we mainly focus on the synchronous aggregation in this paper. After the synchronous aggregation, the updated consensus Q-table $CQ(t)$ is broadcasted to each local Q-learning. Then, the local Q-table is replaced by the consensus Q-table $CQ$. This procedure is shown in Alg.~\ref{Ag1}.

\subsection{Relaxation Optimization}

In this section, we propose an approximation programming, i.e., the Relaxation algorithm, to quickly obtain the suboptimal result. This Relaxation algorithm has a low computation complexity advantage that approximates $P2$ to a convex optimization problem $P3$:
\begin{equation}
\begin{aligned}
\begin{split}
 \textbf{P3:} \qquad &\min_{\bm \varrho} \ \sum\limits_{i=1}^K \left\{{C^{comm}_i(\bm \varrho)  + C^{comp}_i(\bm \varrho) + \delta_i \sum\limits_{h} \ln\varepsilon^h_i(\bm \varrho)}\right\}\\
 \text{s.t.} ,\ \ \ \ \ 
&\text{C1}, \text{C2}, \text{C3}\\
\end{split} 
\end{aligned} 
\label{falarri}
\end{equation}

\noindent where the non-convex part $\max\limits_h \ln\varepsilon_i^h$ of $P2$ is replaced to $\sum\limits_h \ln\varepsilon_i^h$. Since the objective of $P3$ has a gap $\left\{\sum\limits_h \ln\varepsilon^h_i(\bm \varrho) - \max\limits_{h} \ln\varepsilon^h_i(\bm \varrho)\right\}$ with the original problem $P2$, the total cost of $P3$ will be worse (larger) than that of $P2$. However, $\ln\varepsilon^h_i(\bm\varrho)$ is an affine function of $\bm\varrho$ that indicates the linear summation of the affine function $\sum\limits_h \ln\varepsilon^h_i(\bm \varrho)$ is convex. Moreover, $C1$ to $C3$ are linear constraints. Hence, $P3$ is a convex optimization problem. Although the optimal solution of $P3$ is worse than that of $P2$, the computational complexity is greatly simplified because the non-convex problem $P2$ is converted to the convex problem $P3$. It means that the Matlab-based CVX toolbox could solve $P3$ directly.

\section{Performance Evaluation}

To confirm the efficiency of the proposed synchronous federated Q-learning (Sync-FQL), we investigate the asynchronous federated Q-learning (Asyn-FQL), original Q-learning (QL), Greedy algorithm, and Relaxation algorithm to compare with Sync-FQL. The Asyn-FQL method is substantially the same as the Sync-FQL. But the global generation in Asyn-FQL is performed when each local learning has been updated. As for the original Q-learning, it does not have the global aggregation process. The greedy algorithm is devised to always choose the maximum reward $r_i$ for each action selection. The results of the Relaxation algorithm are obtained by directly solving $P3$ via CVX.

\subsection{Performance of Offloading Algorithms}

\begin{figure}
\centering
\subfloat[Total cost of offloading algorithms.]{\label{111111111111}{\includegraphics[width=0.495\linewidth]{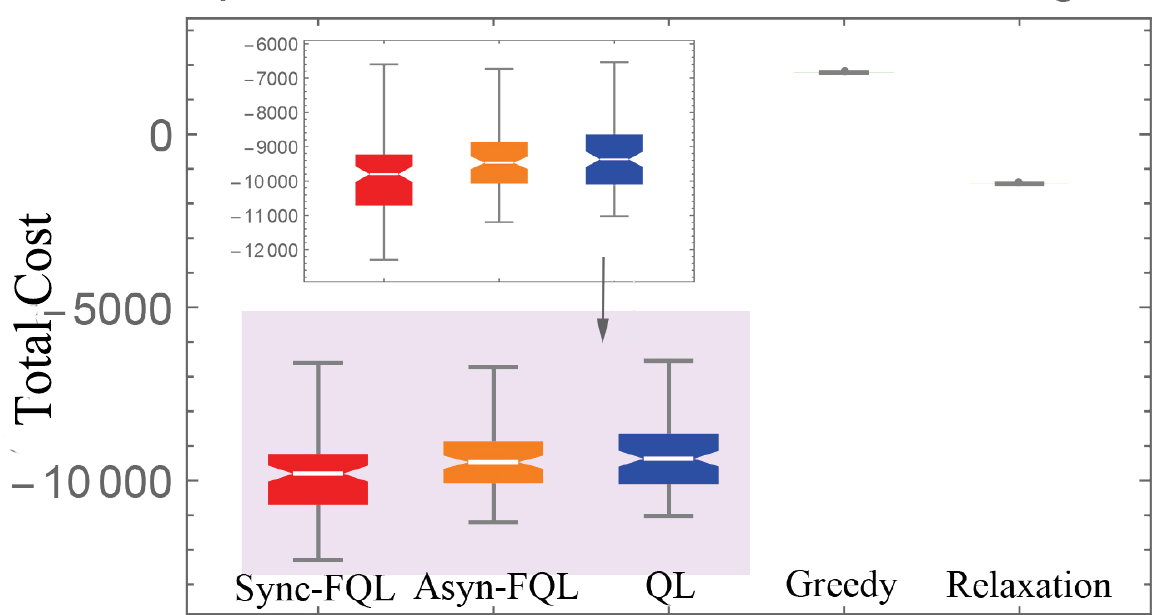}}}  \hfill
\subfloat[Execution time of offloading algorithms.]{\label{222222222222}{\includegraphics[width=0.48\linewidth]{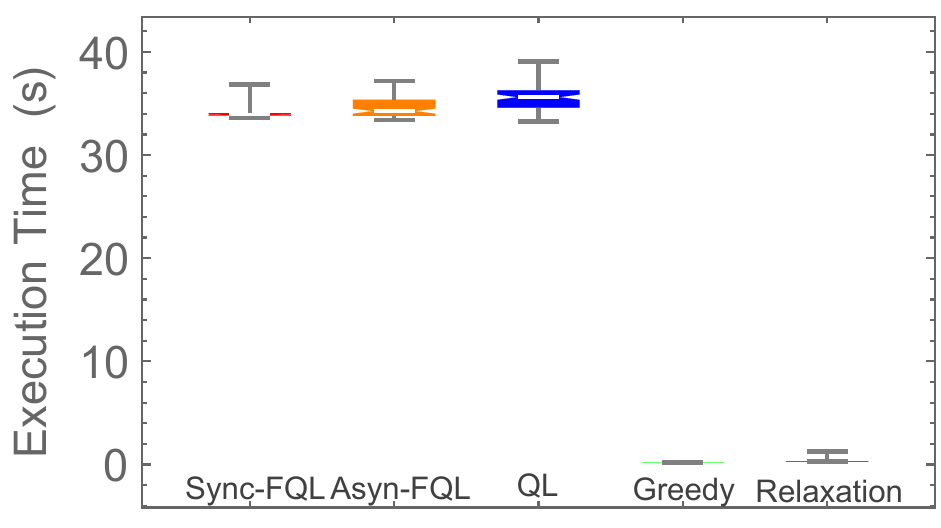}}} \hfill
\caption{Performances of offloading algorithms.}
\label{44444444444444444444444}
\end{figure}

The box-plot of the total cost (i.e., the objective value of $P2$) versus different algorithms is illustrated in Fig.~\ref{111111111111}, where the y-axis represents the total cost of $P2$ of different algorithms. It is collected from 500 simulations. The training period of the learning-based algorithms (i.e., Sync-FQL, Asyn-FQL, and QL) is set to 200 times. There are 5 vehicular tasks generated by each vehicle. The default number of vehicles is set to 5. Without loss of generality,

To simplify the analysis, the priorities of different applications are set equal, i.e., $\bm\delta = [1,1,\dots, 1]$. The computing resource of a service vehicle and the VEC pool are $10^3 Mbps$ and $10^4 Mbps$, respectively. As for the communication bandwidth, the transmission rate of DSRC and C-V2X are $10^3 Mbps$ and $10^4 Mbps$. However, since mmWave communication provides significant large bandwidth compared to the sub-6 GHz technologies, we assume the transmission delay of mmWave communication is negligible. The arrival rates of the tasks are $[20, 70, 30, 80, 8]$, respectively. And the corresponding burstiness measures of different tasks are $[60, 400, 90, 380, 10]$. Moreover, the delay requirement of tasks are $[2, 6, 2.5, 3, 1]$, respectively.

As shown in Fig.~\ref{111111111111}, the proposed Sync-FQL has the lowest $P2$ cost (total cost), and the total cost of Greedy and Relaxation are much worse than that of the learning-based algorithms. The reason is that the proposed federated learning-based algorithm (Sync-FQL and Asyn-FQL) can obtain the learning results of different access technologies via the global aggregation operation. However, due to lacking global aggregation, the original Q-learning cannot share the diverse learning results between different access technologies. In addition, Sync-FQL implements global aggregation when all local learnings have updated. However, Asyn-FQL applies global aggregation when each local learning has updated. Although asynchronous aggregation has proven to be faster than those synchronous, which often results in convergence to poor results. The main potential problem of the asynchronous aggregation is that each local update implements over potentially old parameters.

The execution time of different algorithms is illustrated in Fig.~\ref{222222222222}. In the simulations, each algorithm runs 500 times to obtain stable statistical results. The simulation is implemented by Matlab on a laptop with i5-8300h CPU and 16G RAM. The Greedy and Relaxation algorithms have low execution time because of the simplicity. Regarding the practical vehicular applications, Campolo \textit{et al.} \cite{Campolo2017} pointed out that not all the applications are delay-sensitive. For example, the latency requirements of the vehicular Internet, infotainment, remote diagnostics and management are over 100m. These delay-tolerant applications occupy almost whole vehicular transmission. Once Sync-FQL is trained on a powerful edge server at first, it could meet the requirement of the delay-tolerant applications. As for the delay-sensitive applications, we can apply the proposed Relaxation algorithm rather than the learning-based offloading scheme. Although the performance of the Relaxation is worse than that of learning-based algorithm, the delay-sensitive applications only occupies a small percentage of the wireless communication bandwidths. The overall V2X offloading efficiency will not decline significantly.

\begin{figure}
\subfloat[Performances of different frameworks with light loads.]{\label{wiudhwiuhduwdhuwdhuwhduw}{\includegraphics[width=0.49\linewidth]{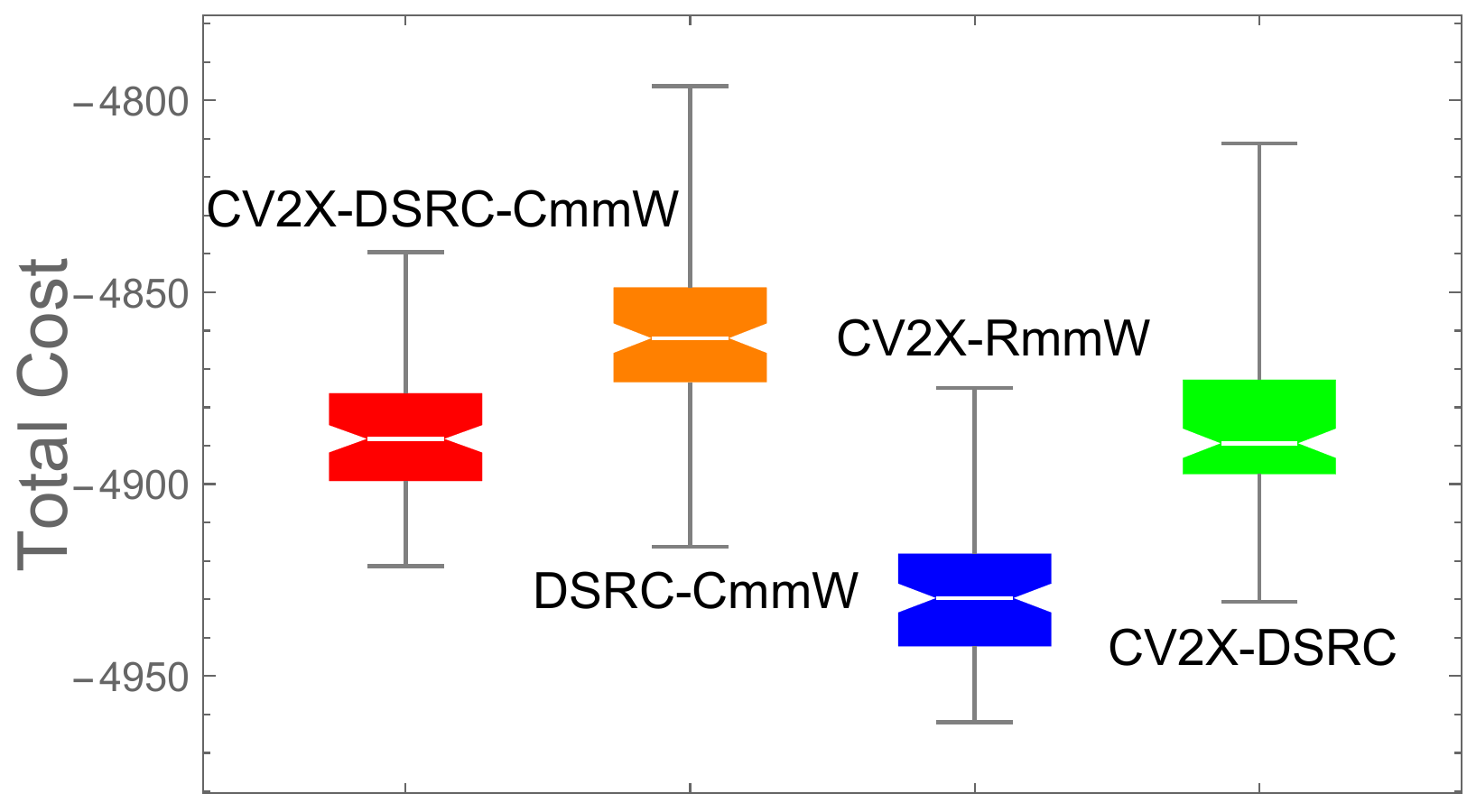}}}  \hfill
\subfloat[Performances of different frameworks with heavy loads.]{\label{djbdjbdbjdbjdbjdbjbdjbjdb}{\includegraphics[width=0.49\linewidth]{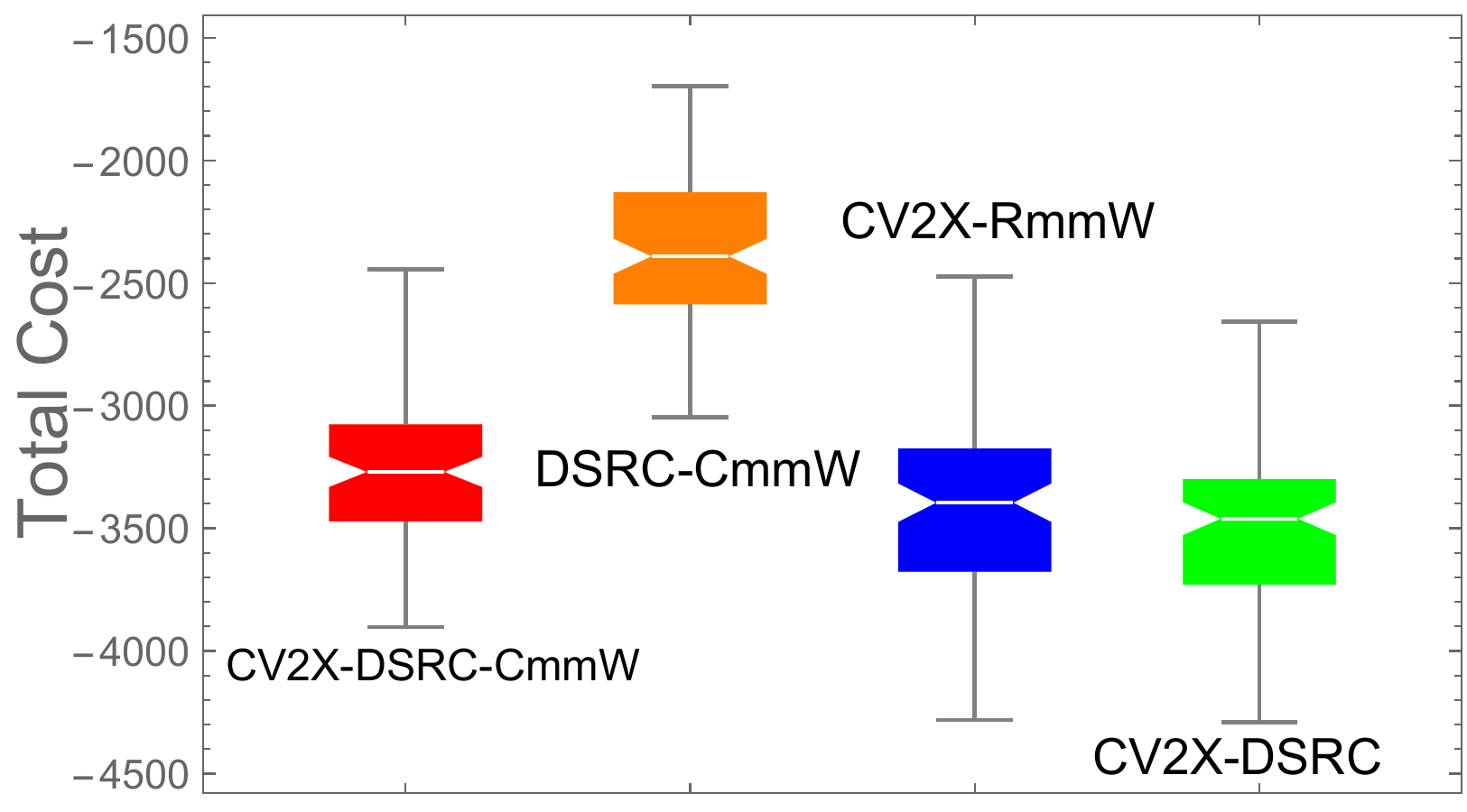}}} \hfill
\caption{Performances of frameworks in different traffic loads.}
\label{jbjbjbjbjbjbjbjbjbjbjbjbjbjbjbjbjbjbjbjb}
\end{figure}

Furthermore, we investigate the performance of heterogeneous offloading frameworks in Fig.~\ref{jbjbjbjbjbjbjbjbjbjbjbjbjbjbjbjbjbjbjbjb}. There are 4 different heterogeneous frameworks: the C-V2X combining with DSRC and CmmW (CV2X-DSRC-CmmW) framework; the DSRC combining with CmmW (DSRC-CmmW) framework; the C-V2X combining with RmmW (CV2X-RmmW) framework; and the C-V2X combining with DSRC (CV2X-DSRC) framework. Since the formation of Eq.(\ref{larrxdddddddddi}) is the same as that of Eq.(\ref{fadkg;jrhgliqghr}) when $o^b_j$ is negligible, the performance of the CV2X-DSRC framework is a good reference for that of the RmmW combing with DSRC (RmmW-DSRC) framework. In the \textit{light traffic loads} scenario, there are 5 categories tasks with the arrival rate $\bm\lambda = [5, 5, 5, 5, 5]$, brustiness measure $\bm o = [2, 2, 2, 2, 2]$, and the delay requirement $\bm T_{max} = [1, 1, 1, 1, 1]$, respectively. While the \textit{heavy traffic loads} of Fig.~\ref{djbdjbdbjdbjdbjdbjbdjbjdb} are set to $\bm\lambda = [100, 100, 100, 100, 100]$ and $\bm o = [50, 50, 50, 50, 50]$ with $\bm T_{max} = [1, 1, 1, 1, 1]$. 

Upon the \textit{light traffic loads} (Fig.~\ref{wiudhwiuhduwdhuwdhuwhduw}), CV2X-RmmW offers the lowest total cost (resource cost and failure probability). In \textit{heavy traffic loads} (Fig.~\ref{djbdjbdbjdbjdbjdbjbdjbjdb}), the performance of CV2X-DSRC is slightly better than that of CV2X-RmmW. Regarding the DSRC-CmmW framework, it has largest total cost. The reason is that CmmW offloading will compete with DSRC offloading for the bandwidth, because the mmWave beam alignment needs the side information transferred by the DSRC channel. In addition, the RmmW offloading can provide better service experience than that of DSRC in the \textit{light traffic loads}. However, the situation converses in \textit{heavy traffic loads}. This is because the mmWave beam alignment requires double transmissions: the RTS-like transmission and CTS-like transmission. Respecting the DSRC offloading, it only has one data transmission. Therefore, the RmmW offloading suffers more deteriorating than that of DSRC upon the heavy traffic.

\begin{remark}
Upon the \textit{light traffic loads}, the C-V2X combining with the RmmW has the lowest total cost (resource cost and the offloading failure probability). However, the C-V2X combining with DSRC offloading has the lowest total cost in the \textit{heavy traffic loads}.
\end{remark}


\begin{figure}
\centering
\subfloat[]{\label{fig_sim_noCrossEntropyScore}{\includegraphics[width=0.5\linewidth]{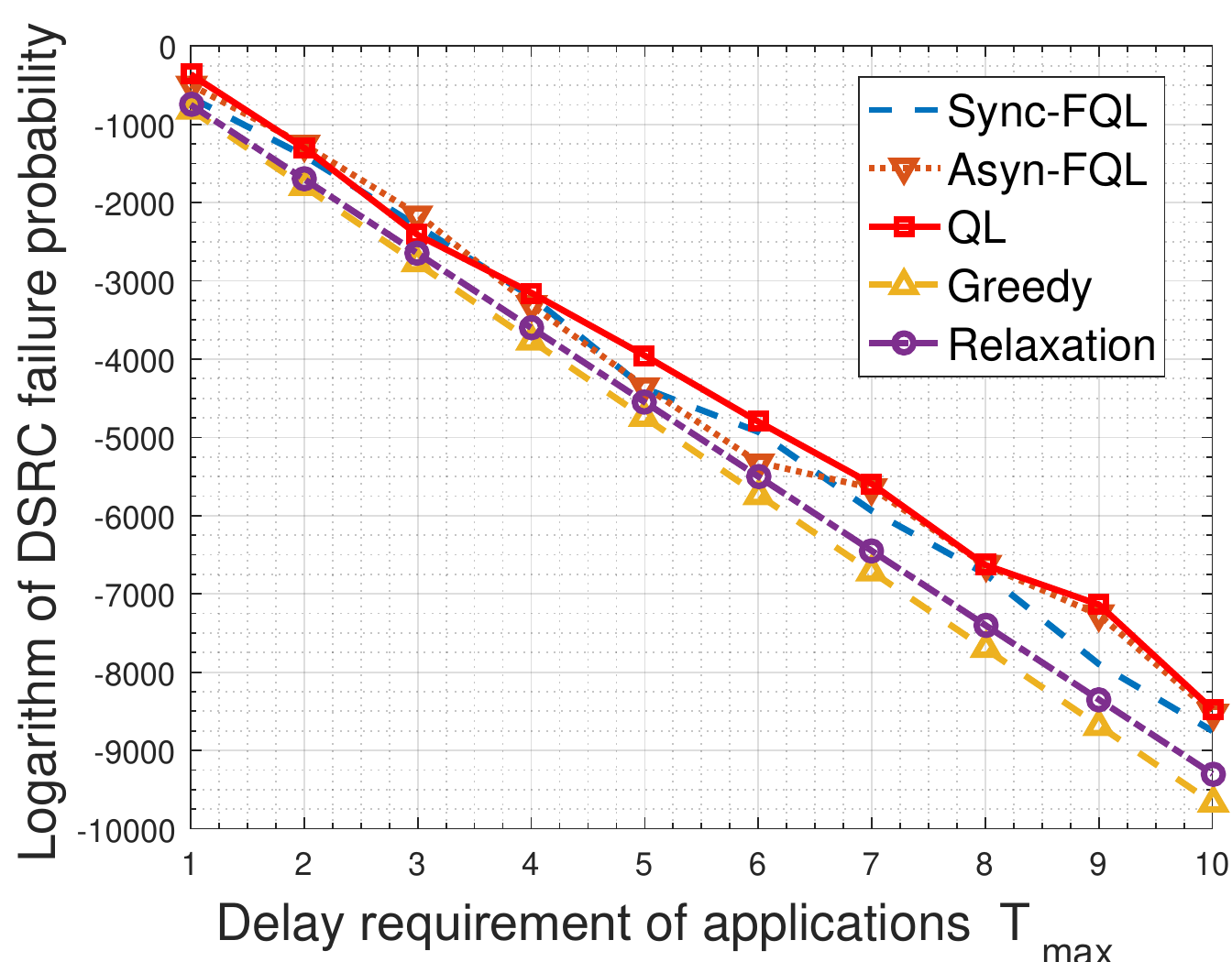}}}\hfill
\subfloat[]{\label{fig_sim_noActionScore}{\includegraphics[width=0.49\linewidth]{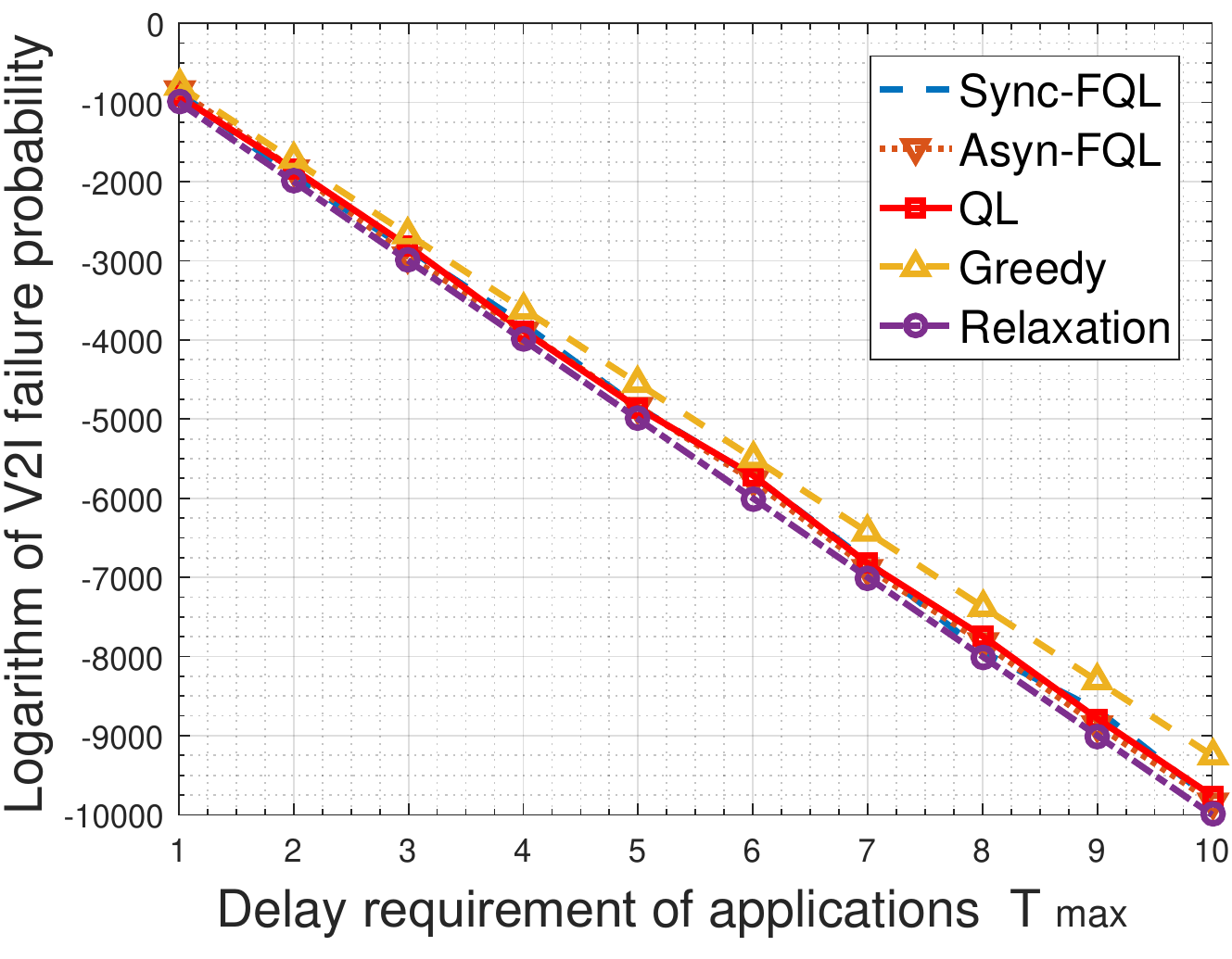}}} \hfill
\subfloat[]{\label{fig_sim_greedyScore}{\includegraphics[width=0.5\linewidth]{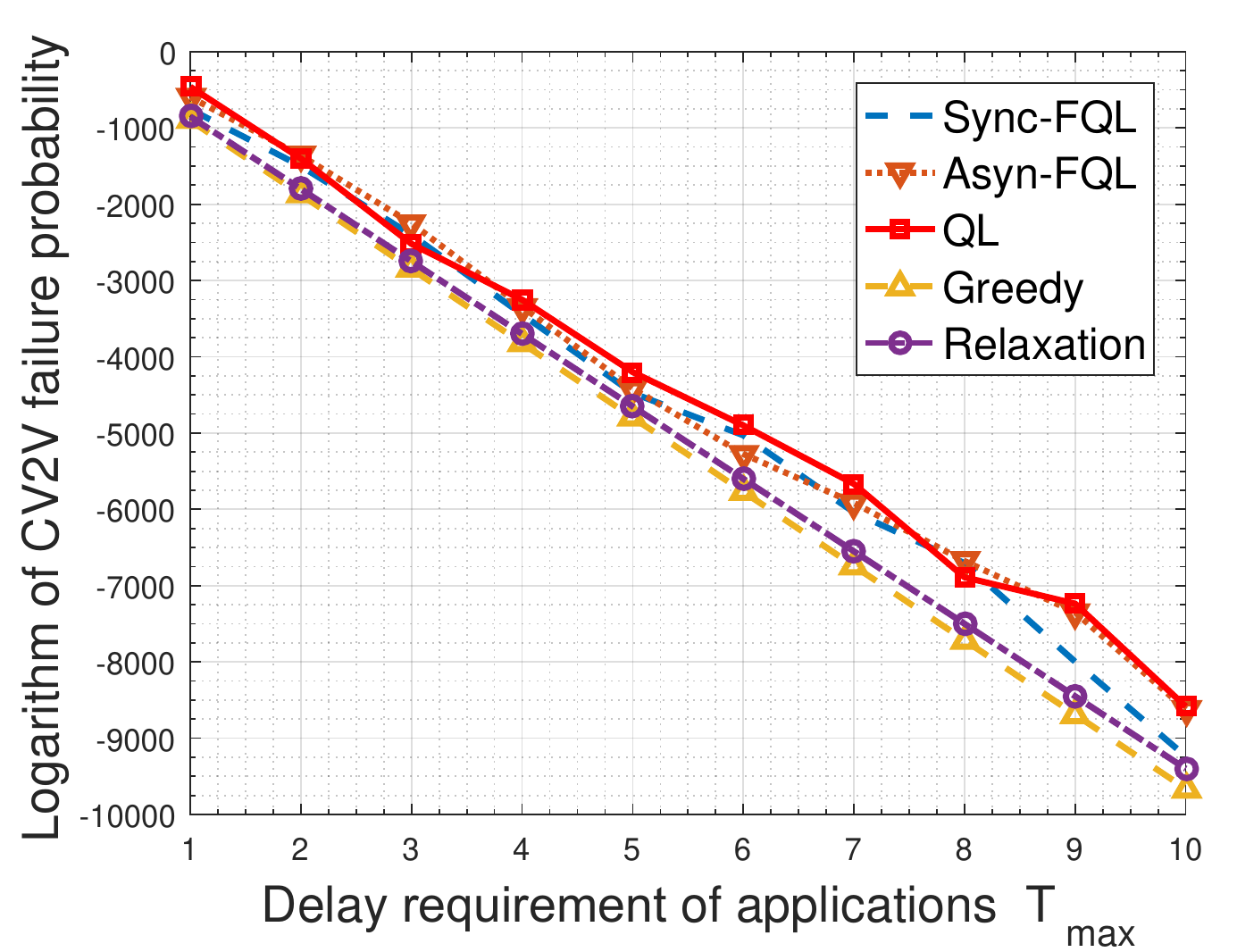}}} \hfill
\subfloat[]{\label{fig_sim_noCrossEntropyScore_0}{\includegraphics[width=0.49\linewidth]{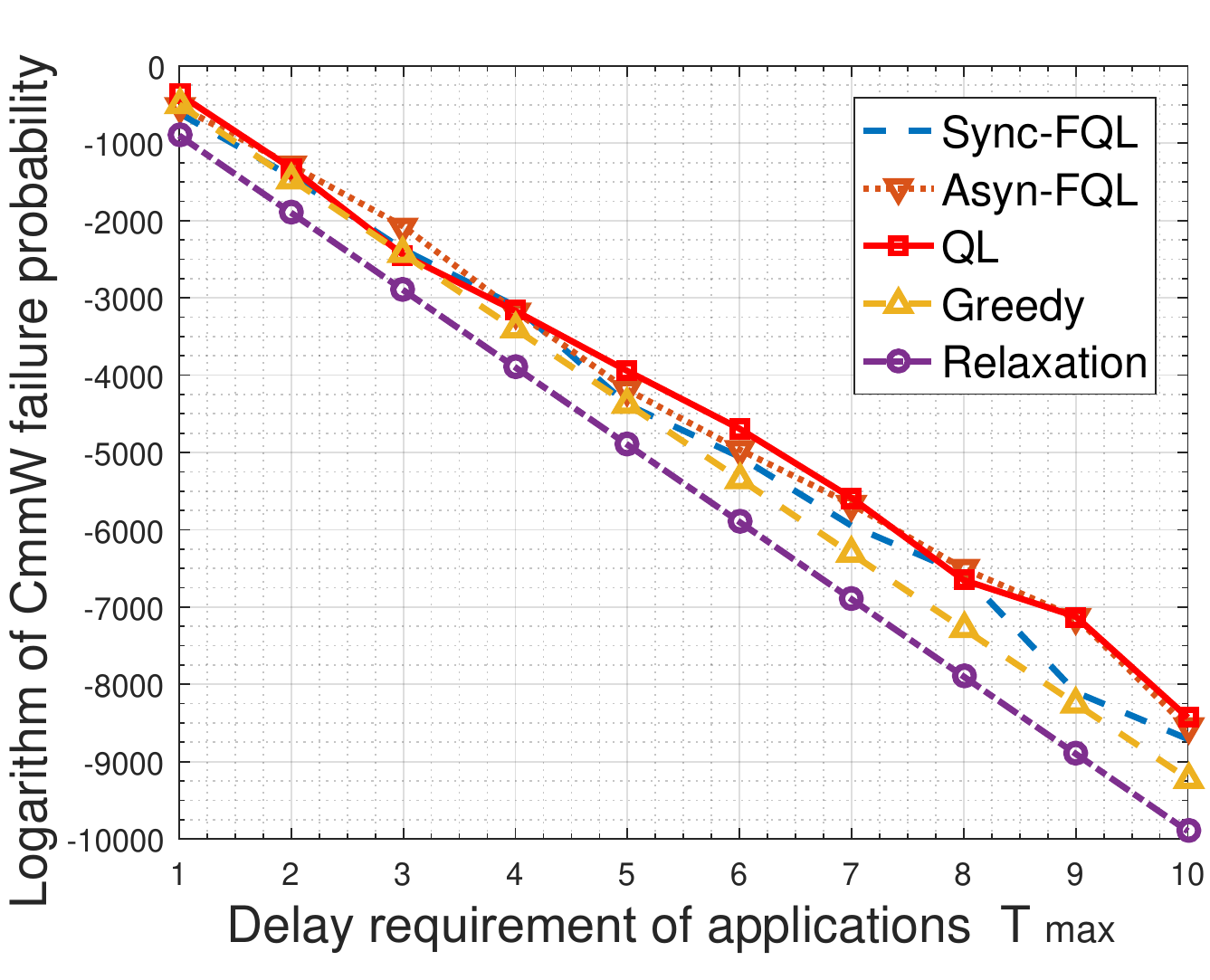}}}\hfill
\subfloat[]{\label{fig_sim_noActionScore_0}{\includegraphics[width=0.5\linewidth]{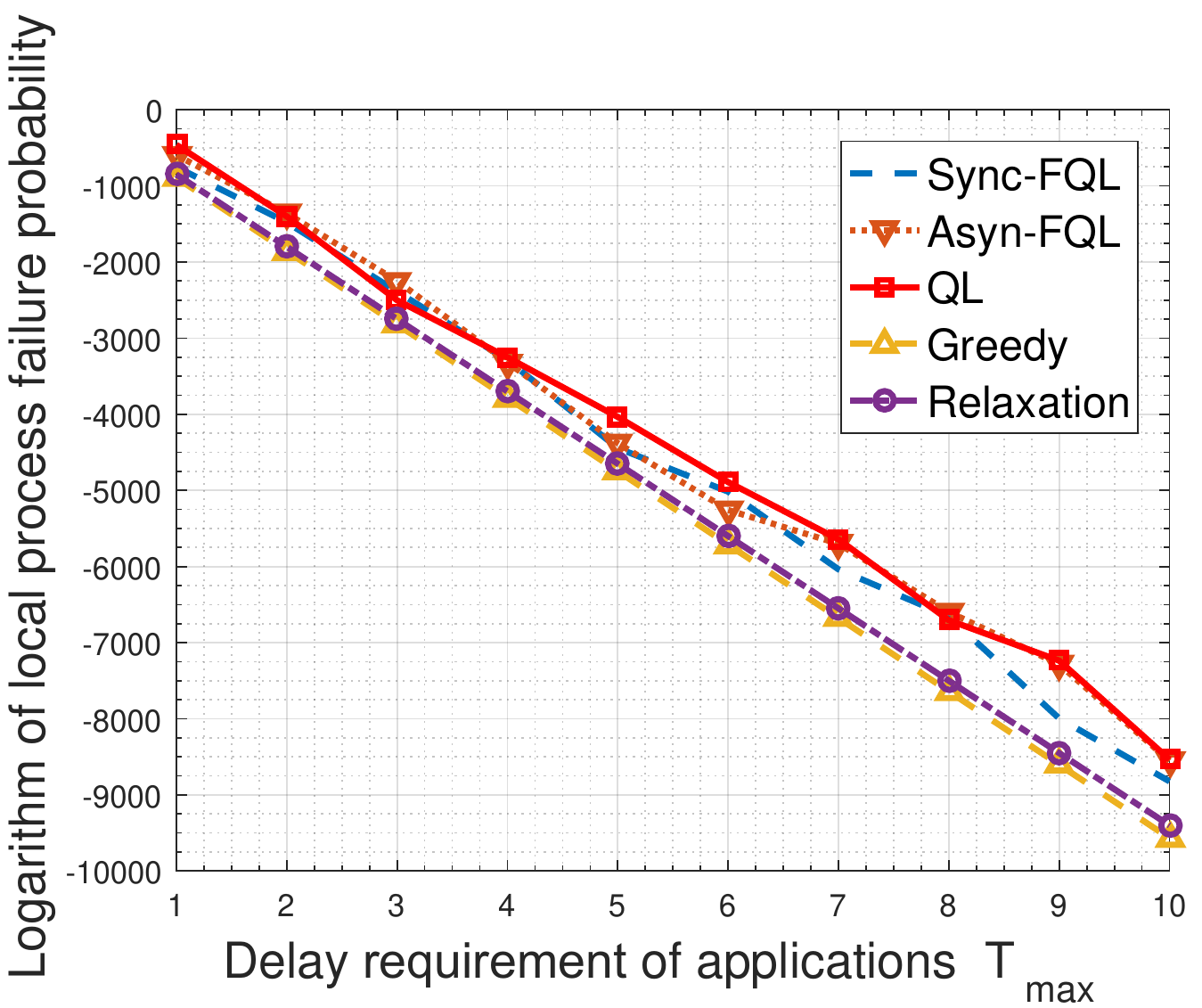}}} \hfill
\subfloat[]{\label{fig_sim_greedyScore_0}{\includegraphics[width=0.49\linewidth]{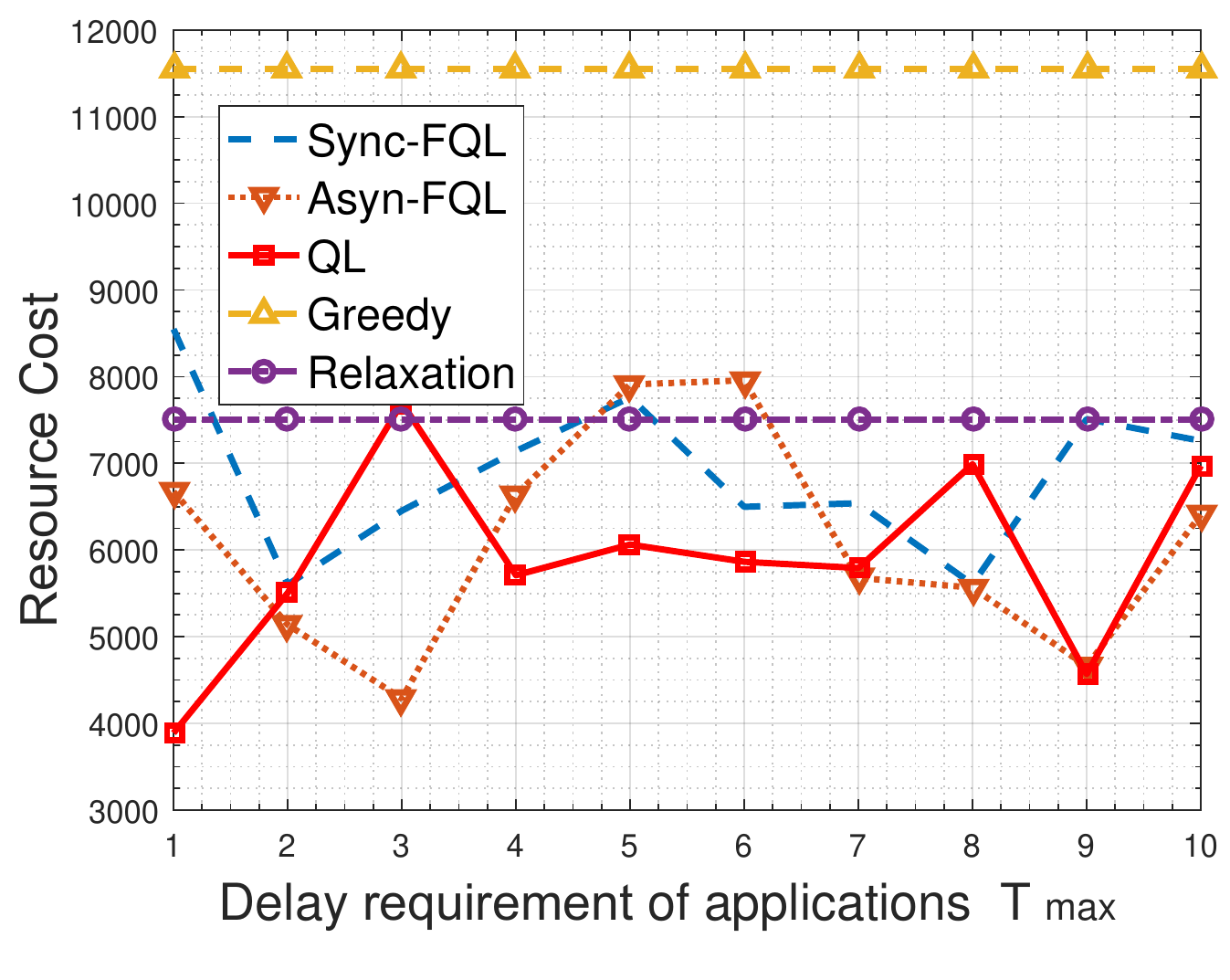}}} \hfill
\caption{(a)-(e) are the logarithm failure probabilities for DSRC, V2I, C-V2V, mmWave, and local processing, respectively. (f) is the resource cost for different algorithms}
\label{fig_sim_scores}
\end{figure}

\begin{figure*}[h]
\centering
\subfloat[Delay upper bound with different arrival rate $\lambda$ under the failure probability $\varepsilon=0.01$ and $o = 100$.]{\label{ssssssss}{\includegraphics[width=0.32\linewidth]{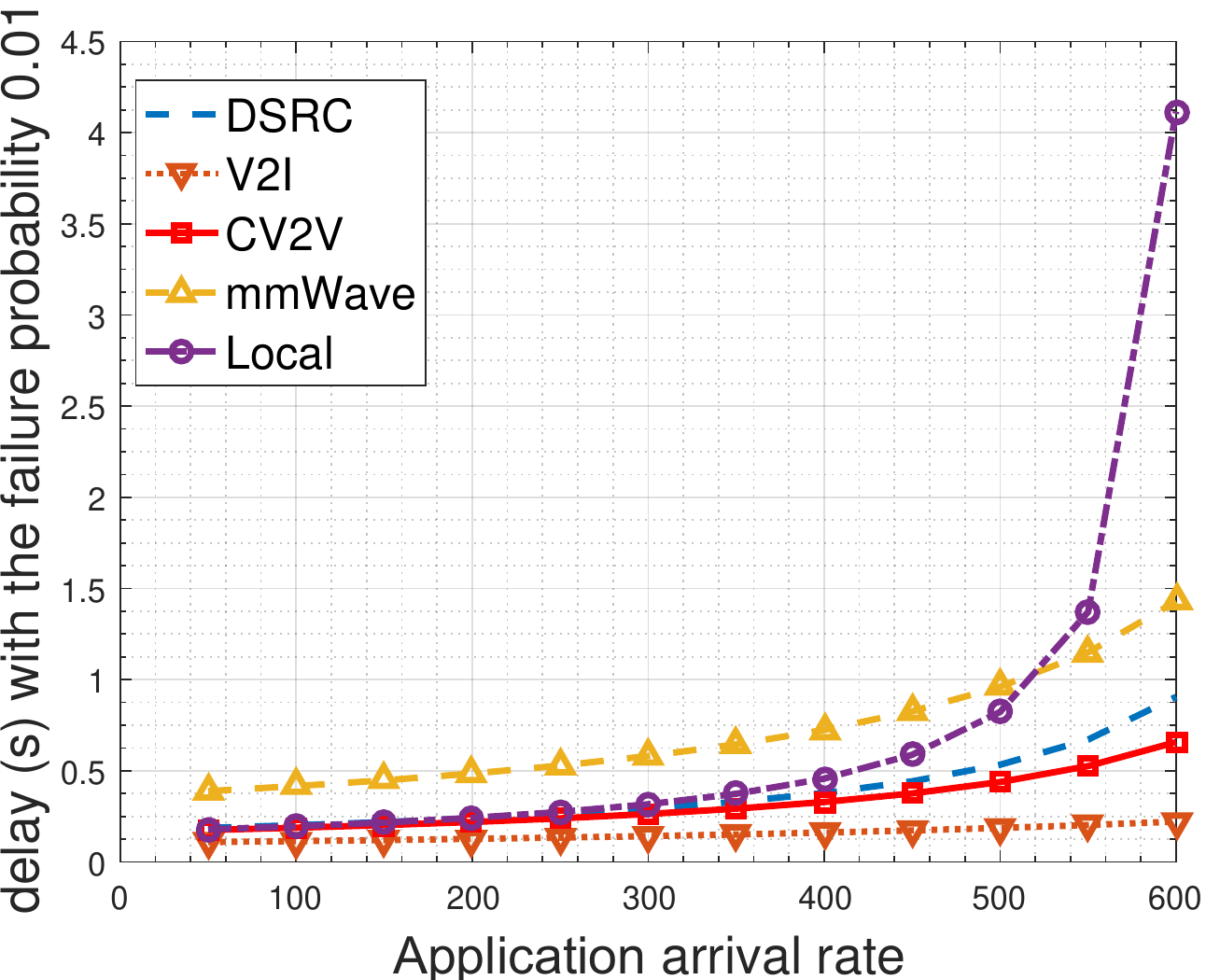}}}  \hfill
\subfloat[Delay upper bound with different burstiness $o$ under the failure probability $\varepsilon=0.01$ and $\lambda = 50$.]{\label{aaaaasasasasa}{\includegraphics[width=0.32\linewidth]{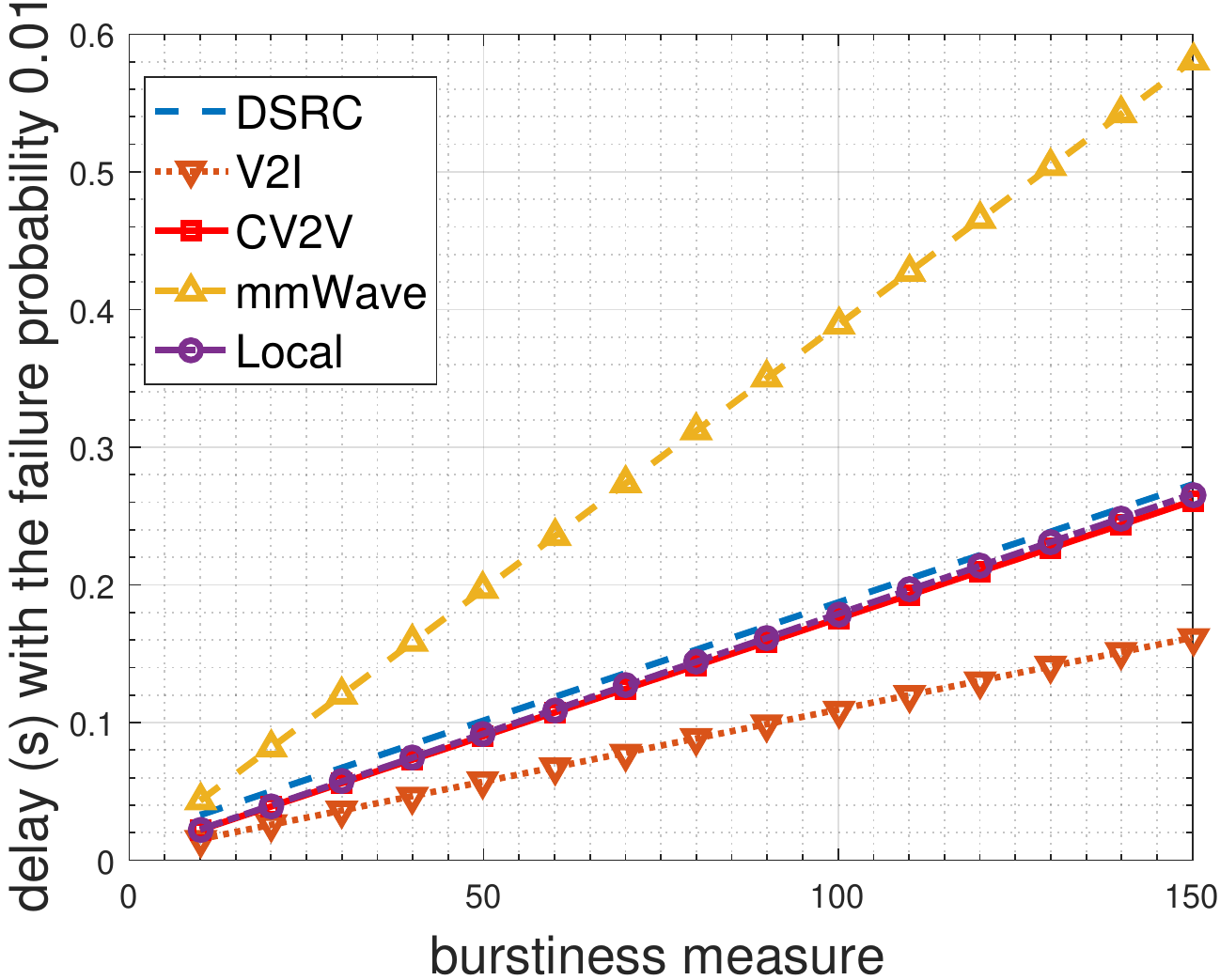}}} \hfill
\subfloat[Logarithm failure probability $\ln \varepsilon$ with different delay requirements.]{\label{ascaljscascoroorrr}{\includegraphics[width=0.33\linewidth]{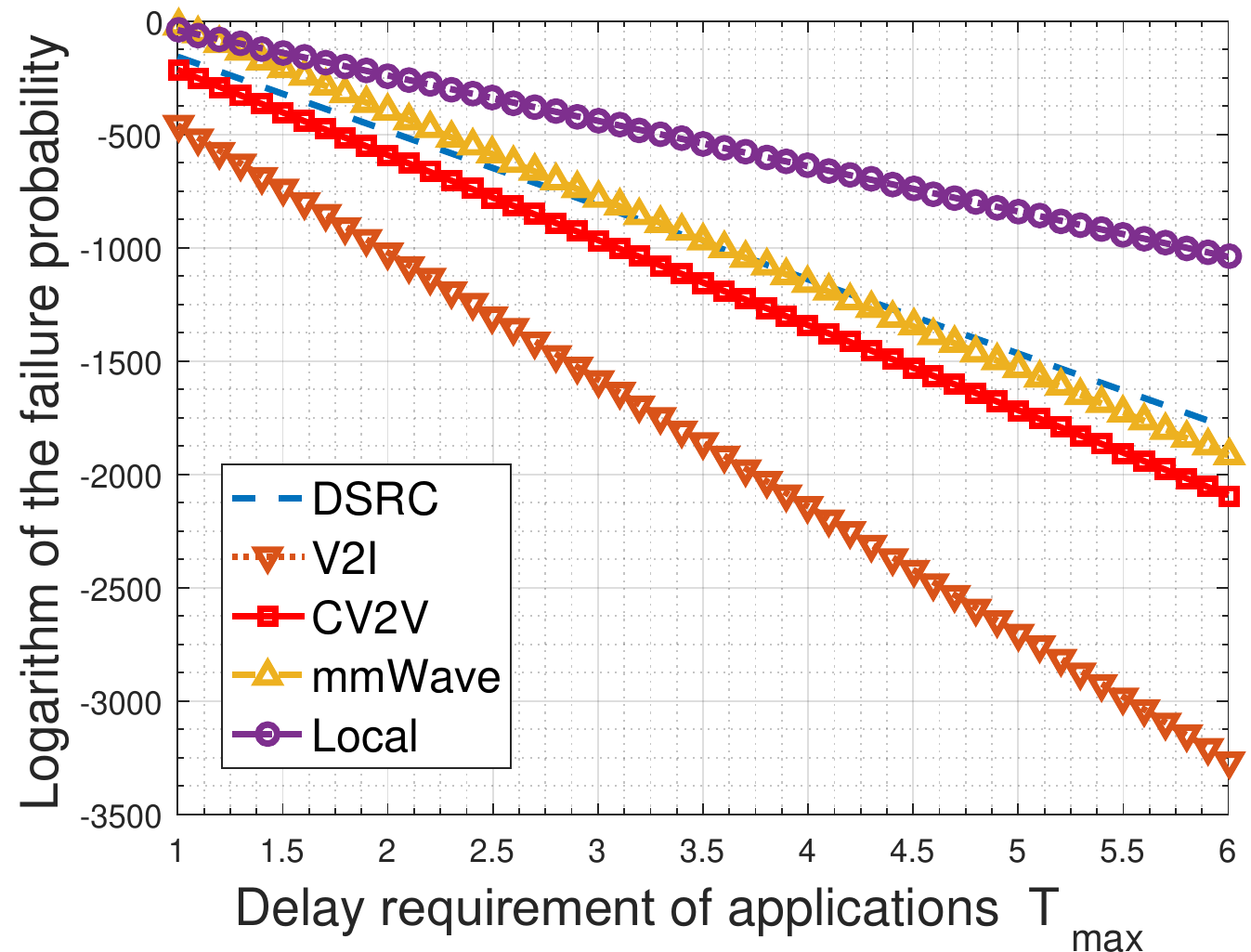}}} \hfill
\caption{Performance of different offloading technologies under various tasks.}
\label{alsdvjnadksjvnlkn}
\end{figure*}

\subsection{Performance of Offloading Technologies}

The failure probability versus delay requirement $T_{max}$ for different V2X technologies depicts in Fig.~\ref{fig_sim_noCrossEntropyScore}-\ref{fig_sim_noActionScore_0}. In \textit{heavy traffic loads}, these offloading failure probabilities are inverse with delay requirement $T_{max}$. Relaxation algorithm has the lowest failure probability. Besides, Fig.~\ref{fig_sim_greedyScore_0} presents the resource cost of different algorithms, where the resource cost is defined as the sum of the communication cost and the computing cost $\sum\limits_{i=1}^K {C^{comm}_i(\bm \varrho)  + C^{comp}_i(\bm \varrho)}$. Because of the inherent stochastic character of learning-based algorithms, the cost curves of learning-based algorithms are fluctuations, while curves of Greedy and Relaxation are constant. And, the resource cost of the Greedy and the Relaxation algorithms are larger than that of learning-based algorithms. Due to the complexity of the $P2$, the Greedy algorithm cannot simultaneously optimize the failure probability and resource cost. As for the Relaxation algorithm, since the objective of $P3$ includes the sum of all failure probabilities, the weight of the optimization tends to the failure probability rather than the resource cost. This is the reason that the total cost of Greedy and Relaxation are worse than others in Fig.~\ref{44444444444444444444444}.

\subsection{Performance with Traffic Loads}

Fig.~\ref{alsdvjnadksjvnlkn} verifies the impact of different traffic loads on the offloading. Fig.~\ref{ssssssss} shows the delay upper bounds of different V2X technologies increased with the arrival rate $\lambda$ of tasks. In this simulation, the acceptable failure probability is set to 0.01, and the burstiness is set to $100 Mbp$. When the arrival rate excesses $500 Mpbs$, the delay upper bound of the local process grows sharply. It reveals that offloading the heavy workloads to others (i.e., vehicles or VEC servers) instead of processing locally, is an efficient approach to reduce the offloading delay. Fig.~\ref{aaaaasasasasa} illustrates that the delay upper bounds of V2X technologies raised with the burstiness measure $o$. In addition, the delay upper bound of mmWave increases faster than that of other V2X technologies. That means mmWave offloading is not suitable for heavy burstiness traffic. However, in Fig.~\ref{ascaljscascoroorrr}, the C-V2I offloading has the lowest delay upper bound. Thus, a task has the large burstiness trait that prefers to C-V2I offloading.

Fig.~\ref{ascaljscascoroorrr} illustrates the failure probability decreased with the rising of delay requirements $T_{max}$ of tasks. MmWave communication declines slowest than the other offloading technologies, which demonstrates that the mmWave offloading is not the best choice for the delay-tolerant traffic. However, in Fig.~\ref{ascaljscascoroorrr}, C-V2V offloading (or RmmW offloading) has the lowest failure probability. Therefore, the delay-tolerant task prefers to C-V2V offloading for the lowest failure probability.

\begin{remark}
Vehicles should offload the large arrival rate traffic to others; the CmmW offloading is not proper to handle the tasks with the heavy burstiness measure. And, a delay-tolerant task prefers to employ the C-V2V offloading for the lowest failure probability. 
\end{remark}


\section{Conclusion}

In this paper, we have studied the task offloading in the heterogeneous VEC system with various V2X communications and edge servers. To minimize the offloading failure probability and resource cost, we first derive the upper bound of offloading delay with offloading failure probability through the stochastic network calculus. These upper bounds can be used as an indicator to guarantee the quality of services for automotive tasks. Hereafter, we devote to efficient offloading in the presence of task offloading failure (transmission failed and computing interrupted), and propose an intelligent offloading scheme to ensure offloading reliability while reducing resource cost. Simulations show that the C-V2X combining with RmmW offloading and the C-V2X combining with DSRC offloading have the best performance in the light and heavy traffic loads, respectively. Upon the traffic load with a high arrival rate, vehicles should offload the high arrival rate traffic to others. However, when tasks become more bursty, vehicles should avoid using CmmW communication. Besides, a delay-tolerant application should utilize the C-V2V offloading for the low offloading failure probability.


\appendices

\footnotesize
\bibliography{biblio}

\end{document}